\documentclass{elsarticle}

\usepackage{amsmath,amssymb}
\usepackage{color}
\usepackage{float}
\usepackage{array}
\usepackage{rotating}
\usepackage{algpseudocode}
\usepackage{algorithm}
\usepackage{subfigure}
\usepackage{soul}
\mathchardef\mhyphen="2D
\usepackage[normalem]{ulem}
\usepackage{multirow}
\usepackage{bm}
\usepackage{dcolumn}
\usepackage{amsfonts}
\usepackage[colorlinks=true,linkcolor=blue,citecolor=blue,urlcolor=blue]{hyperref}

\biboptions{sort&compress}

\newcommand{\Tmatrix}[0]{\mathbb{T}_\mathbf{q}^\mu}

\def\QE{\textsc{Quantum ESPRESSO}}

\def\turboMagnon{\texttt{turboMagnon}\,}

\usepackage[normalem]{ulem}
\newcommand{\editor}[2]{%
  \expandafter\newcommand\csname #1note\endcsname[1]{%
    \textcolor{#2}{(\textbf{#1:} ##1)}}%
  \expandafter\newcommand\csname #1\endcsname[1]{%
    \textcolor{#2}{##1}}%
  \expandafter\newcommand\csname #1cancel\endcsname[1]{%
    \textcolor{#2}{\sout{##1}}}%
  \expandafter\newcommand\csname #1change\endcsname[2]{%
    \textcolor{#2}{\sout{##1} ##2}}%
  \newenvironment{#1text}{\color{#2}}{\color{black}}
}

\definecolor{tangerine}{rgb}{0.944,0.522,0}
\definecolor{fuchsia}{rgb}{0.792, 0.173, 0.729}
\definecolor{random}{rgb}{0.399,0.400,0.800}

\begin{document}

\begin{frontmatter}

\title{\turboMagnon - A code for the simulation of spin-wave spectra using the Liouville-Lanczos approach to time-dependent density-functional perturbation theory} 
\author[paris]{Tommaso Gorni\corref{author}}
\author[sissa]{Oscar Baseggio}
\author[sissa]{Pietro Delugas}
\author[sissa]{Stefano Baroni}
\author[epfl]{Iurii Timrov\corref{author2}}
\cortext[author] {Corresponding author. \textit{e-mail address:} gornitom@gmail.com}
\cortext[author2] {Corresponding author. \textit{e-mail address:} iurii.timrov@epfl.ch}
\address[paris]{LPEM, ESPCI Paris, PSL Research University, CNRS, Sorbonne Universit\'e, 75005 Paris France, European Union}
\address[sissa]{SISSA -- Scuola Internazionale Superiore di Studi Avanzati, Trieste, Italy, European Union} 
\address[epfl]{{T}heory and Simulation of Materials (THEOS) and National Centre for Computational Design and Discovery of Novel Materials (MARVEL), \'{E}cole Polytechnique F\'{e}d\'{e}rale de Lausanne, CH-1015 Lausanne, Switzerland}


\begin{abstract}
We introduce \turboMagnon, an implementation of the Liouville-Lanczos approach to linearized time-dependent density-functional theory, designed to simulate spin-wave spectra in solid-state materials. The code is based on the noncollinear spin-polarized framework and the self-consistent inclusion of spin-orbit coupling that allow to model complex magnetic excitations. The spin susceptibility matrix is computed using the Lanczos recursion algorithm that is implemented in two flavors - the non-Hermitian and the pseudo-Hermitian one. \turboMagnon is open-source software distributed under the terms of the GPL as a component of \QE. As with other components, \turboMagnon is optimized to run on massively parallel architectures using native mathematical libraries (LAPACK and FFTW) and a hierarchy of custom parallelization layers built on top of MPI. The effectiveness of the code is showcased by computing magnon dispersions for the CrI$_3$ monolayer, and the importance of the spin-orbit coupling is discussed.
\end{abstract}

\end{frontmatter}
{\bf PROGRAM SUMMARY}

\begin{small}
  \noindent
  {\em Program title:} \turboMagnon \\
  {\em Licensing provisions:} GNU General Public License V 2.0  \\
  {\em Programming language:} Fortran 95                         \\
  {\em Computer:} Any computer architecture                      \\
  {\em External routines:} \turboMagnon is a tightly integrated component of the
  \QE\, distribution and requires the standard libraries linked by it: BLAS, LAPACK, FFTW, MPI.\\
  {\em Nature of problem:} Calculation of the spin-wave spectra for solid-state materials.  
  \\
  {\em Solution method:} The spin susceptibility matrix of a periodic system is expressed in terms of the resolvent of its Liouvillian superoperator within time-dependent density-functional perturbation theory. It is calculated using non-Hermitian or pseudo-Hermitian variants of the Lanczos recursion scheme, whose implementation does not require the calculation of empty electronic states. Norm-conserving pseudopotentials are used in conjunction with plane-wave basis sets and periodic boundary conditions. Relativistic effects (spin-orbit coupling) can be included in calculations.
  \\
  {\em Additional comments including restrictions and unusual features:} Linear-response regime only. Collinear spin-polarized formalism is not supported, only noncollinear spin-polarized case can be used. Adiabatic LSDA exchange-correlation functionals only (no GGA, no meta-GGA, no hybrid functionals, no Hubbard $U$, etc.). No ultrasoft and projector-augmented-wave pseudopotentials. No symmetry.\\ 
  Empty states are not used, nor even calculated. Three Lanczos recursions give access to the whole spectrum of magnetic excitations at fixed transferred momentum; one Lanczos recursion is enough for a specific orientation of the magnetic field and the ground-state magnetization of the system. \\
  The distribution file of this program can be downloaded from the \QE\, website: http://www.quantum-espresso.org/, and the development version of this program can be downloaded via Git from the GitLab website:\\ https://gitlab.com/QEF/q-e . Interactions with end users of the \turboMagnon code happen via a mailing-list forum of \QE: https://www.quantum-espresso.org/forum. Documentation of the \turboMagnon code is tightly coupled with the code and is done via standard code comments.
\end{small}

\section{Introduction}
\label{sec:intro}

The characterization of magnetic excitations at the atomistic level has become possible in the last 50 years thanks to the development and continuous refinement of magnetic spectroscopies, most notably  inelastic neutron scattering spectroscopy (INSS) for bulk materials~\cite{Mook:1973}, spin-polarized electron energy loss spectroscopy (SPEELS) and inelastic scanning tunneling spectroscopy for thin films~\cite{Qin:2015, Hirjibehedin:2006}.
Moreover, a promising recent development is represented by resonant inelastic x-ray scattering (RIXS) spectroscopy, which has been shown able to probe magnetic excitations both in bulk~\cite{Chaix:2018,Brookes:2020,Lebert:2020} and thin crystals~\cite{Pellicciari:2021,Pelliciari:2021aa}.
The interpretation of these spectroscopies usually relies on model Hamiltonians, possibly in conjunction with ground-state \emph{ab initio} calculations to fit the parameters appearing therein~\cite{Costa:2010, Bergman:2010, Zakeri:2012, Zakeri:2017}. 
These treatments, however, are formally justified only in the limit of localized magnetic moments and cannot capture most of the features of itinerant magnetic systems.
Even in the former case, they face difficulties in complying with the growing number of parameters when more exchange paths are at play or in the presence of magnetic anisotropies, such as the ones induced by spin-orbit coupling.
An alternative route is represented by a fully \emph{ab~initio} treatment of spin dynamics, which can be pursued via the computation of the dynamical spin susceptibility from either time-dependent density-functional theory (TDDFT)~\cite{Savrasov:1998, Lounis:2011, Buczek:2011b, Rousseau:2012, dosSantosDias:2015, Wysocki:2017, Cao:2018, TancogneDejean:2020, Skovhus:2021} or many-body perturbation theory (MBPT)~\cite{Aryasetiawan:1999, Karlsson:2000, Kotani:2008, Sasioglu:2010, Muller:2016}. 
Both these methods treat charge and spin fluctuations on an equal footing in a self-consistent manner and they are formally exact, though in practice they rely on different approximations and have different computational requirements. 
TDDFT is numerically way less demanding than MBPT, particularly when adopting the adiabatic local spin density approximation (ALSDA), which often results in a good compromise between computational cost and accuracy~\cite{Runge:1984, Marques:2012} and has in fact been widely adopted for modeling magnetic excitations.
Previous attempts to compute magnon dispersion relations from linear-response TDDFT were based on either the solution of the time-dependent Sternheimer equation \cite{Savrasov:1998, Cao:2018} or of the Dyson equation for the spin susceptibility, starting from the independent-electron spin and charge susceptibilities \cite{Lounis:2011, Buczek:2011b, Rousseau:2012, dosSantosDias:2015, Wysocki:2017, Skovhus:2021}.
In all these approaches the linear-response problem must be solved for every individual value of the excitation frequency, which is one of the main computational bottlenecks to be addressed and overcome in this work. 
We remark here that, more recently, magnetic excitations were computed  beyond the linear response regime from TDDFT via a real-time propagation technique~\cite{TancogneDejean:2020}.

In this paper we introduce a computer code, named \turboMagnon, which implements the Liouville-Lanczos (LL) approach to INSS and SPEELS spectroscopies within time-dependent density-functional perturbation theory (TDDFpT)~\cite{Baroni:2012, Rocca:2008, Timrov:2013}. 
This allows us to treat the dynamical spin-fluctuation response of magnetic systems in a fully noncollinear framework, and thus model their spin-wave excitation spectra including spin-orbit-coupling effects entirely from first principles. 
By using techniques borrowed from static density-functional perturbation theory (DFpT)~\cite{Baroni:1987, Baroni:2001} and similarly to the method of Ref.~\cite{Cao:2018}, our method avoids computing independent-particle susceptibilities, and thus does not require computationally expensive and slowly converging sums over empty states. 
At variance with previous studies, in the case of adiabatic TDDFT kernels our method also avoids repeated linear-response calculations for each individual excitation frequency, by using a recursive Lanczos algorithm to obtain a tridiagonal matrix representation of the Liouvillian superoperator. 
The actual spectrum is then computed in an inexpensive post-processing step for any desired frequency. This allows us to obtain the full spectrum of magnetic excitations (both magnons and Stoner excitations) in any frequency range with three Lanczos chains per excitation wave-number in the most general case, where the full $3\times 3$ spin susceptibility matrix is needed. 
\turboMagnon has a similar structure as the \texttt{turboTDDFT} and \texttt{turboEELS} codes used to compute absorption spectra in molecular systems~\cite{Malcioglu:2011, Ge:2014} and electron energy loss spectra in periodic solids~\cite{Timrov:2013, Timrov:2015, Timrov:2017, Motornyi:2020}. 
All these two codes share a large number of linear response routines which are gathered in the \texttt{LR\_Modules} repository of \QE{}. 
\turboMagnon is distributed under the terms of the GPL license \cite{GPL}, as a component of the \QE\, suite of open-source codes based on plane-wave basis sets, pseudopotentials, and using periodic boundary conditions~\cite{Giannozzi:2009, Giannozzi:2017, Giannozzi:2020}. 
 
This paper is organized as follows. 
In Sec.~\ref{sec:theory} we provide a theoretical background for the LL approach to TDDFpT for magnetic excitations. 
In Sec.~\ref{sec:Description_of_software_components} we describe \turboMagnon and the calculation workflow for computing magnons. 
In Sec.~\ref{sec:installation_and_parallelization} we provide the instructions for installing \turboMagnon and discuss various levels of parallelization that can be used.
In Sec.~\ref{sec:benchmarking} we show how to use \turboMagnon for computing magnon dispersions in the CrI$_3$ monolayer in the noncollinear framework with or without spin-orbit coupling, how to check the sum rules, and also we discuss the scaling of the code. 
Finally, Sec.~\ref{sec:Conclusions} contains conclusions and perspectives for future work.


\section{Theory}
\label{sec:theory}

\subsection{Statement of the problem}
\label{sec:Statement_of_the_problem}

In INSS experiments a neutron beam with wave-vector $\mathbf{k}_i$ and energy $E_i$ impinges
on the target sample. Due to inelastic scattering, the outgoing neutron will be characterized
by the wave-vector $\mathbf{k}_f = \mathbf{k}_i - \mathbf{q}$ and energy
$E_f = E_i - \hbar \omega$, where $\hbar\mathbf{q}$ and $\hbar \omega$ are the momentum and energy
transferred to the sample, respectively. In the first Born approximation~\cite{Halpern:1939,Blume:1963}, the double-differential
cross section corresponding to magnetic excitations of electrons can be written in the compact form as:
\begin{equation}
\frac{d^2 \sigma}{d\Omega d\omega} =
\frac{\hbar}{\pi}\left(\frac{g_n e}{2\hbar}\right)^2\frac{k_f}{k_i} \, S(\mathbf{q},\mathbf{\omega}) \,,
\label{eq:cross_section}
\end{equation}
where
\begin{equation}
S(\mathbf{q},\mathbf{\omega}) = -{\rm Im}\,{\rm Tr} \bigg[ {\boldsymbol P}^{\perp}(\mathbf{q}) \,
{\boldsymbol \chi}(\mathbf{q},\mathbf{q}; \omega) \bigg] \,.
\label{eq:S_def}
\end{equation}
Here, $-e$ and $g_n \approx 3.826$ are the electron charge and the neutron $g$-factor, respectively,
${\boldsymbol P}^{\perp}(\mathbf{q})$ is the $3 \times 3$ matrix, $P_{\alpha\beta}^{\perp}(\mathbf{q}) =
\delta_{\alpha\beta} - q_{\alpha}q_{\beta}/q^2$ (with $\alpha , \beta = x,y,z$), which is a projector on to the plane perpendicular to the direction of $\mathbf{q}$, and ${\boldsymbol \chi}(\mathbf{q},\mathbf{q};\omega)$ is the $3 \times 3$ spin susceptibility matrix~(see, e.g., Sec.~5.7 in Ref.~\cite{Jones:1985}). The poles of $S(\mathbf{q},\mathbf{\omega})$ occur at frequencies of magnons and Stoner excitations. This quantity is accessible from linear-response theory, and in the following we will show how it can be computed efficiently using the LL approach to TDDFpT. Hereafter, Hartree atomic units will be used, and the formalism is presented for insulating systems while the generalization to metals can be found in Appendix~B of Ref.~\cite{Gorni:2018} and in Ref.~\cite{Gorni:2016}.


\subsection{Time-dependent density-functional perturbation theory}

Magnetic excitations in solids can be modeled using the time-dependent Pauli-type Kohn-Sham (KS) equations of TDDFT that read:
\begin{equation}
i \frac{\partial \Psi_{n,\mathbf{k}}(\mathbf{r},t)}{\partial t} = \hat{H}(t) \Psi_{n,\mathbf{k}}(\mathbf{r},t) ,
\label{eq:TDKS}
\end{equation}
where $n$ and $\mathbf{k}$ are the electronic band index and the crystal momentum, respectively, $\Psi_{n,\mathbf{k}}(\mathbf{r},t)$ are the two-component time-dependent KS spinor wave functions, and $\hat{H}(t)$ is the time-dependent Hamiltonian operator of the system~\cite{Notations}. It is often convenient to use perturbation theory to first order within TDDFT (i.e. TDDFpT) what is also known as the \textit{linear-response} TDDFT. Moreover, Eq.~\eqref{eq:TDKS} is often solved in the frequency domain rather than in the time domain. Therefore, it can be shown that but linearizing Eq.~\eqref{eq:TDKS} and by making a Fourier transform from the time domain to the frequency domain we can obtain the following set of the so-called \textit{resonant} and \textit{anti-resonant} Pauli-type linear-response KS (Sternheimer) equations~\cite{Gorni:2016, Gorni:2018}:
\begin{align}
\bigl( \hat{H}^{\circ}_\mathbf{k+q} - \varepsilon^\circ_{n,\mathbf{k}} - \omega \bigr) \,
\tilde{U}'_{n,\mathbf{k+q}}(\mathbf{r},\omega) & + \, \hat{P}_{\mathbf{k+q}} \,
\hat{\tilde{V}}^{\prime}_{\mathrm{HXC},\mathbf{q}}(\omega) \,
U^\circ_{n,\mathbf{k}}(\mathbf{r}) 
= - \hat{P}_{\mathbf{k+q}} \,
\hat{\tilde{V}}^{\prime}_{\mathrm{ext},\mathbf{q}}(\omega)
\, U^\circ_{n,\mathbf{k}}(\mathbf{r}) \,,
\label{eq:KS_eq_resonant_projected_q} 
\\
\bigl( \hat{H}^{\circ +}_\mathbf{k+q} - \varepsilon^\circ_{n,-\mathbf{k}} + \omega \bigr) \,
\hat{\mathrm{T}} \tilde{U}'_{n,-\mathbf{k}-\mathbf{q}}(\mathbf{r},-\omega) & + \, \hat{P}^{+}_{\mathbf{k+q}} \,
\hat{\tilde{V}}^{\prime +}_{\mathrm{HXC},\mathbf{q}}(\omega) \,
\hat{\mathrm{T}} U^\circ_{n,-\mathbf{k}}(\mathbf{r}) 
= - \hat{P}^{+}_{\mathbf{k+q}} \,
\hat{\tilde{V}}^{\prime +}_{\mathrm{ext},\mathbf{q}}(\omega) \,
\hat{\mathrm{T}} U^\circ_{n,-\mathbf{k}}(\mathbf{r}) \,.
\label{eq:KS_eq_antiresonant_projected_q}
\end{align}
All the quantities appearing in the equations above are defined in the following. We want to note in passing that a step-by-step derivation of the equations \eqref{eq:KS_eq_resonant_projected_q} and \eqref{eq:KS_eq_antiresonant_projected_q} starting from the ground-state problem is presented in Ref.~\cite{Gorni:2018}; here we are presenting briefly the final results. In equations \eqref{eq:KS_eq_resonant_projected_q} and \eqref{eq:KS_eq_antiresonant_projected_q}
we are dealing with the lattice-periodic quantum-mechanical operators and functions. In particular, we have used the following definitions for the ground-state and linear-response KS spinor wave functions: $\Psi^\circ_{n,\mathbf{k}}(\mathbf{r}) = e^{i\mathbf{k}\cdot\mathbf{r}} \, U^\circ_{n,\mathbf{k}}(\mathbf{r}) / \sqrt{N_\mathbf{k}}$ and $\tilde{\Psi}'_{n,\mathbf{k+q}}(\mathbf{r},\omega) = e^{i(\mathbf{k+q})\cdot\mathbf{r}} \,
\tilde{U}'_{n,\mathbf{k+q}}(\mathbf{r},\omega) / \sqrt{N_\mathbf{k}}$, where $\mathbf{q}$ is the wavevector of the perturbation, $N_\mathbf{k}$ is the number of $\mathbf{k}$ points in the Brillouin zone (BZ), $U^\circ_{n,\mathbf{k}}(\mathbf{r})$ and $\tilde{U}'_{n,\mathbf{k+q}}(\mathbf{r},\omega)$ are the lattice-periodic functions. Equations~\eqref{eq:KS_eq_resonant_projected_q} and \eqref{eq:KS_eq_antiresonant_projected_q} must be solved for a specific perturbing potential, that in the cases of spin waves is defined as $\hat{\tilde{V}}^{\prime}_{\mathrm{ext},\mathbf{q}}(\omega) =
- \mu_\mathrm{B} \, {\boldsymbol \sigma} \cdot
\hat{\tilde{\boldsymbol b}}'_{\mathrm{ext},\mathbf{q}}(\omega)$, where $\mu_\mathrm{B}$ is the Bohr magneton, ${\boldsymbol \sigma} = (\sigma_x, \sigma_y, \sigma_z)$ is the vector of Pauli matrices, and $\hat{\tilde{\boldsymbol b}}'_{\mathrm{ext},\mathbf{q}}(\omega)$ is the lattice-periodic operator of the external magnetic field~\cite{Baroni:2001}.
Equation~\eqref{eq:KS_eq_antiresonant_projected_q} was obtained from Eq.~\eqref{eq:KS_eq_resonant_projected_q} by changing the sign of $\omega$, $\mathbf{k}$, and $\mathbf{q}$, and by applying the time-reversal operator $\hat{\mathrm{T}} = i \sigma_y \hat{K}$, where $\hat{K}$ is the complex-conjugation operator. The operator $\hat{H}^{\circ}_\mathbf{k}$ has the real-space representation  $H^\circ_\mathbf{k}(\mathbf{r},\mathbf{r}') = e^{-i\mathbf{k}\cdot\mathbf{r}}H^\circ(\mathbf{r},\mathbf{r}')e^{i\mathbf{k}\cdot\mathbf{r}'}$, 
and we defined for convenience
$\hat{H}^{\circ +}_\mathbf{k+q} \equiv \hat{\rm T} \hat{H}^{\circ}_\mathbf{-k-q}  \hat{\rm T}^{-1}$, where $\hat{H}^\circ$ is the Hamiltonian operator of the unperturbed system. The definition of $\hat{H}^\circ_\mathbf{k+q}$ is as follows~\cite{Gorni:2018}:
\begin{equation}
\hat{H}^\circ_\mathbf{k+q} =
\sigma^\circ \biggl[ -\frac{1}{2} \left[\nabla + i (\mathbf{k+q})\right]^2 
+ \, \hat{v}^\circ_\mathrm{loc} + \hat{v}^\circ_\mathrm{H} + 
\hat{v}^\circ_\mathrm{XC} \biggr] + \hat{V}^\circ_\mathrm{NL, \mathbf{k+q}} -
\mu_\mathrm{B} \, {\boldsymbol \sigma} \cdot \hat{{\boldsymbol b}}^\circ_\mathrm{XC} \,,
\label{eq:H0_k_plus_q}
\end{equation}
where $\sigma^\circ$ is the unit $2 \times 2$ matrix, the first term in Eq.~\eqref{eq:H0_k_plus_q} is the kinetic energy, $\hat{v}^\circ_\mathrm{loc}$ is the local part of the pseudopotential (PP)~\cite{Pot_notation}, $\hat{V}^\circ_\mathrm{NL, \mathbf{k+q}}$ is the $2 \times 2$ matrix which is the nonlocal part of the PP containing the scalar-relativistic and spin-orbit coupling (SOC) potentials~\cite{Kleinman:1980, Bachelet:1982a, Bachelet:1982b, Hemstreet:1993}, $v^\circ_\mathrm{H}(\mathbf{r}) = \int \frac{n^\circ(\mathbf{r}')}{|\mathbf{r} - \mathbf{r}'|} \, d\mathbf{r}'$ is the Hartree ground-state potential, $v^\circ_\mathrm{XC}(\mathbf{r}) = \frac{\delta E_\mathrm{XC}[n,{\boldsymbol m}]}{\delta n}$ and ${\boldsymbol b}^\circ_\mathrm{XC}(\mathbf{r}) = - \frac{\delta E_\mathrm{XC}[n,{\boldsymbol m}]}{\delta {\boldsymbol m}}$ are respectively the scalar and magnetic parts of the exchange-correlation (XC) ground-state potential [the functional derivatives defining $v^\circ_\mathrm{XC}(\mathbf{r})$ and ${\boldsymbol b}^\circ_\mathrm{XC}(\mathbf{r})$ are evaluated at ground-state charge density $n=n^\circ(\mathbf{r})$ and magnetization density ${\boldsymbol m} = {\boldsymbol m}^\circ(\mathbf{r})$]. In Eqs.~\eqref{eq:KS_eq_resonant_projected_q} and \eqref{eq:KS_eq_antiresonant_projected_q}, $\varepsilon^\circ_{n,\mathbf{k}}$ and $\varepsilon^\circ_{n,-\mathbf{k}}$ are the ground-state KS energies. It is important to remark that the inclusion of SOC when noncollinear magnetism is explicitly contemplated is trivial, because SOC is time-reversal invariant~\cite{Gorni:2016}; in practice this means that $\hat{\mathrm{T}} \hat{V}^\circ_\mathrm{NL, \mathbf{-k-q}} = \hat{V}^\circ_\mathrm{NL, \mathbf{k+q}} \hat{\mathrm{T}}$ on the left-hand side of Eq.~\eqref{eq:KS_eq_antiresonant_projected_q}, and thus $\hat{H}^{\circ +}_\mathbf{k+q}$ differs from $\hat{H}^\circ_\mathbf{k+q}$ only by the opposite sign in the last term of Eq.~\eqref{eq:H0_k_plus_q}.

It is important to note that here we are using a number of approximations. Firstly, the external perturbing potential $\hat{\tilde{V}}^{\prime}_{\mathrm{ext},\mathbf{q}}(\omega)$ (defined above) contains only a coupling between the spin angular momentum and the magnetic field (i.e. the spin Zeeman term), while we have neglected the coupling between the orbital angular momentum and the magnetic field (i.e. the orbital Zeeman term). This is motivated by the fact that we are considering a regime of a weak (vanishing) external magnetic field. It was shown that in this case for first-row transition-metal elements the orbital Zeeman term is very small compared to the spin Zeeman term and hence it can be neglected (see e.g. Refs.~\cite{dosSantosDias:2015, Ceresoli:2010}). Secondly, we neglected the diamagnetic term (it is proportional to the second power of the magnetic field amplitude). The diamagnetic term is generally much smaller than the spin (plus orbital) Zeeman term and in particular in the case of weak magnetic fields (which is the case here)~\cite{Ashcroft:1976}. Finally, we follow Ref.~\cite{DalCorso:2010} and use the common approximation: only interaction terms up to $1/c^2$ are considered, where $c$ is the speed of light. For this reason we neglected the interaction between the spin-orbit coupling and the magnetic field that is a higher-order term proportional to $1/c^3$ (see e.g. Ref.~\cite{Ceresoli:2010}).


Now let us describe the linear-response potentials in Eqs.~\eqref{eq:KS_eq_resonant_projected_q} and \eqref{eq:KS_eq_antiresonant_projected_q}. $\hat{\tilde{V}}^{\prime}_{\mathrm{HXC},\mathbf{q}}(\omega)$ is the monochromatic $\mathbf{q}$ component of the response Hartree and XC (HXC) potential, which reads:
\begin{equation}
\hat{\tilde{V}}^{\prime}_{\mathrm{HXC},\mathbf{q}}(\omega) = \sigma^\circ \, \hat{\tilde{v}}'_{\mathrm{H},\mathbf{q}}(\omega) + \sigma^\circ \, \hat{\tilde{v}}'_{\mathrm{XC},\mathbf{q}}(\omega) - \mu_\mathrm{B} \, {\boldsymbol \sigma} \cdot \hat{\tilde{\boldsymbol b}}'_{\mathrm{XC},\mathbf{q}}(\omega) \,,
\label{eq:V_HXC_resp_q}
\end{equation}
where
\begin{equation}
\tilde{v}'_{\mathrm{H},\mathbf{q}}(\mathbf{r},\omega) =
\int \frac{\tilde{n}'_\mathbf{q}(\mathbf{r}',\omega)}{|\mathbf{r}-\mathbf{r}'|} \,
e^{-i\mathbf{q}\cdot(\mathbf{r}-\mathbf{r}')} \, d\mathbf{r}'
\label{eq:v_H_q}
\end{equation}
is the response Hartree potential in the coordinate representation, 
while $\hat{\tilde{v}}'_{\mathrm{XC},\mathbf{q}}(\omega)$ and 
$\hat{\tilde{\boldsymbol b}}'_{\mathrm{XC},\mathbf{q}}(\omega)$ are the response scalar and magnetic XC potentials, respectively, which in the coordinate representation within ALSDA read~\cite{Baroni:2001, Note:notation_Vxc}:
  \begin{align}
    \tilde{v}'_{\mathrm{XC},\mathbf{q}}(\mathbf{r},\omega) & = \frac{\partial v_\mathrm{XC}}{\partial n} \biggr|_{n^\circ,{\boldsymbol m}^\circ} \tilde{n}'_\mathbf{q}(\mathbf{r},\omega) + \, \frac{\partial v_\mathrm{XC}}{\partial {\boldsymbol m}} \biggr|_{n^\circ,{\boldsymbol m}^\circ} \tilde{\boldsymbol m}'_\mathbf{q}(\mathbf{r},\omega) \,, \label{eq:v_XC_q} \\
    \tilde{\boldsymbol b}'_{\mathrm{XC},\mathbf{q}}(\mathbf{r},\omega) & = \frac{\partial {\boldsymbol b}_\mathrm{XC}}{\partial n} \biggr|_{n^\circ,{\boldsymbol m}^\circ} \tilde{n}'_\mathbf{q}(\mathbf{r},\omega) + \, \frac{\partial {\boldsymbol b}_\mathrm{XC}}{\partial {\boldsymbol m}} \biggr|_{n^\circ,{\boldsymbol m}^\circ} \tilde{\boldsymbol m}'_\mathbf{q}(\mathbf{r},\omega) \,. \label{eq:b_XC_q}
  \end{align}
From Eqs.~\eqref{eq:v_XC_q} and \eqref{eq:b_XC_q} we can see that there are mixed scalar and magnetic responses of $v_\mathrm{XC}$ and ${\boldsymbol b}_\mathrm{XC}$, which are coupled in a self-consistent way. As will be seen in the following, this allows us to compute the spin susceptibility directly by avoiding calculations of charge-charge responses and cross-terms spin-charge responses. The response potentials in Eqs.~\eqref{eq:v_H_q} -- \eqref{eq:b_XC_q} are expressed in terms of the monochromatic $\mathbf{q}$ components of the response charge and magnetization densities, which read:
\begin{align}
\tilde{n}'_\mathbf{q}(\mathbf{r},\omega) & = 
\frac{1}{N_\mathbf{k}} \sum_{n, \mathbf{k}} \bigg[ f_{n,\mathbf{k}} \, U^{\circ \dagger}_{n,\mathbf{k}}(\mathbf{r}) \, \tilde{U}'_{n,\mathbf{k+q}}(\mathbf{r},\omega) 
+ f_{n,-\mathbf{k}} \left(\hat{\rm T}U^\circ_{n,-\mathbf{k}}(\mathbf{r}) \right)^{\dag} \hat{\mathrm{T}}\tilde{U}'_{n,-\mathbf{k}-\mathbf{q}}(\mathbf{r},-\omega) \bigg] \,,
\label{eq:charge_dens_resp_q} 
\\
\tilde{\boldsymbol m}'_\mathbf{q}(\mathbf{r},\omega) & = 
\frac{\mu_\mathrm{B}}{N_\mathbf{k}} \sum_{n, \mathbf{k}}
\bigg[ f_{n,\mathbf{k}} \, U^{\circ \dagger}_{n,\mathbf{k}}(\mathbf{r}) \, {\boldsymbol \sigma} \, \tilde{U}'_{n,\mathbf{k+q}}(\mathbf{r},\omega) 
- f_{n,-\mathbf{k}} \left(\hat{\rm T}U^\circ_{n,-\mathbf{k}}(\mathbf{r}) \right)^{\dag} \, {\boldsymbol \sigma} \, \hat{\mathrm{T}}\tilde{U}'_{n,-\mathbf{k}-\mathbf{q}}(\mathbf{r},-\omega) \bigg] \,,
\label{eq:spin_dens_resp_q}
\end{align}
and satisfy the following relations~\cite{Note:density_property}:
$\tilde{n}^{\prime *}_{-\mathbf{q}}(\mathbf{r},-\omega) = \tilde{n}'_\mathbf{q}(\mathbf{r},\omega)$ and $\tilde{\boldsymbol m}^{\prime *}_{-\mathbf{q}}(\mathbf{r},-\omega) = \tilde{\boldsymbol m}'_\mathbf{q}(\mathbf{r},\omega)$.
In Eqs.~\eqref{eq:charge_dens_resp_q} and \eqref{eq:spin_dens_resp_q}, $f_{n,\mathbf{k}}$ and $f_{n,-\mathbf{k}}$ are the occupation factors that are equal to 1 for occupied states and to 0 for empty states at zero temperature.
Using the aforementioned properties of the response densities it is easy to show that 
$ \hat{\tilde{V}}^{\prime +}_{\mathrm{HXC},\mathbf{q}}(\omega) \equiv
\hat{\rm T} \, \hat{\tilde{V}}^{\prime}_{\mathrm{HXC},-\mathbf{q}}(-\omega) \hat{\rm T}^{-1}$
is the operator of Eq.~\eqref{eq:V_HXC_resp_q} with the opposite sign in the response magnetic XC potential. The same applies for $\hat{\tilde{V}}^{\prime +}_{\mathrm{ext},\mathbf{q}}(\omega) 
\equiv \hat{\rm T} \, \hat{\tilde{V}}^{\prime}_{\mathrm{ext},-\mathbf{q}}(-\omega) \hat{\rm T}^{-1}$, which is the external perturbing potential with a reversed direction of the magnetic field. Lastly, the operators $\hat{P}_{\mathbf{k+q}}$ and
$\hat{P}^{+}_{\mathbf{k+q}}$, appearing in Eqs.~\eqref{eq:KS_eq_resonant_projected_q} and \eqref{eq:KS_eq_antiresonant_projected_q}, respectively, are the projectors on to the empty-states manifold, and in the coordinate representation they read:
  \begin{align}
    P_{\mathbf{k+q}}(\mathbf{r},\mathbf{r}') & = \delta(\mathbf{r}-\mathbf{r}') -
\sum_{m} f_{m,\mathbf{k}+\mathbf{q}}\, U^\circ_{m,\mathbf{k+q}}(\mathbf{r}) \, U^{\circ \dagger}_{m,\mathbf{k+q}}(\mathbf{r}') \,, \label{eq:P_minus_q2} \\
    P^{+}_{\mathbf{k+q}}(\mathbf{r},\mathbf{r}') & =
\hat{\rm T} \, P_{-\mathbf{k}-\mathbf{q}}(\mathbf{r},\mathbf{r}')  \, \hat{\rm T}^{-1} \nonumber \\
    &  =  \delta(\mathbf{r}-\mathbf{r}') - \sum_{m} f_{m,-\mathbf{k}-\mathbf{q}}\left( \hat{\mathrm{T}} U^\circ_{m,-\mathbf{k}-\mathbf{q}}(\mathbf{r}) \right) \left( \hat{\mathrm{T}} U^{\circ}_{m,-\mathbf{k}-\mathbf{q}}(\mathbf{r}') \right)^\dagger \,. \label{eq:P_plus_q2}
  \end{align}
We stress that these projectors are expressed in terms of the ground-state spinors $U^{\circ}_{m,\mathbf{k}+\mathbf{q}}$ and 
$\hat{\mathrm{T}} U^{\circ}_{m,-\mathbf{k}-\mathbf{q}}$, respectively, which in turn refer to the occupied-states manifold (similarly to the static DFpT~\cite{Baroni:1987,Baroni:2001}) and thus avoiding the computationally expensive summations over empty states. Finally, it is important to stress that the self-consistent solution of the coupled equations~\eqref{eq:KS_eq_resonant_projected_q} and \eqref{eq:KS_eq_antiresonant_projected_q} is done for a fixed external perturbation with a fixed wavevector $\mathbf{q}$, which thus does not require the mixing of responses to different $\mathbf{q}$-specific perturbations which greatly simplifies the solution of the problem.

We note that the Sternheimer equations~\eqref{eq:KS_eq_resonant_projected_q} and \eqref{eq:KS_eq_antiresonant_projected_q} can be solved directly by using e.g. the conjugate-gradient algorithm~\cite{Cao:2018}.
However, this requires solving a separate self-consistent problem for each frequency $\omega$, so increasing the computational cost very rapidly for very dense $\omega$ grids.
In the next section, we present an alternative way to solve these equations using the LL approach, which allows us to bypass the aforementioned bottleneck via the use of an effective recursive Lanczos algorithm.

We recall that the formalism presented above is based on ALSDA. The extension to generalized-gradient approximation (GGA), especially in the noncollinear framework, is quite involved and not considered here. In this respect it is useful to mention Ref.~\cite{Singh:2019} where it was shown that adiabatic GGA generally worsens the spin-excitation spectra by overestimating the magnon energies and suppressing the intensity of spin waves. The extension to meta-GGA functionals is even more challenging, in particular because of the known numerical stability issues even when performing ground-state calculations~\cite{Lehtola:2022}, not to mention the difficulties in generalizing meta-GGA to (time-dependent) linear-response theory. Moreover, e.g. the recently-proposed SCAN meta-GGA functional~\cite{Sun:2015} exhibits some potential limitations in describing magnetic system~\cite{Ekholm:2018, Tran:2020}. As a note of caution, when attempting to generalize TDDFpT to GGA and meta-GGA functionals one has to pay special attention on to whether the zero-torque theorem is still satisfied. Finally, as possible future extensions of the current TDDFpT formalism it would be interesting and important to perform generalizations to hybrid and Hubbard functionals which allow to better describe the localized $d$ and $f$ electrons by alleviating large self-interaction errors for these states~\cite{Sasioglu:2010, Skovhus:2022, Skovhus:2022b}.


\subsection{Quantum Liouville equation and spin susceptibility matrix}
\label{sec:quantum_Liouville_eq}

The resonant and anti-resonant linear-response KS equations~\eqref{eq:KS_eq_resonant_projected_q} and \eqref{eq:KS_eq_antiresonant_projected_q} can be equivalently expressed in terms of the quantum Liouville equation for the $2 \times 2$ response spin-charge density matrix operator $\hat{\tilde{\rho}}_{\mathbf q}'(\omega)$~\cite{Timrov:2013}:
\begin{equation}
( \omega - \hat{\mathcal{L}}_\mathbf{q} ) \cdot \hat{\tilde{\rho}}_{\mathbf q}'(\omega)
= [\hat{\tilde{V}}^{\prime}_\mathrm{ext,\mathbf{q}}(\omega), \hat{\rho}^\circ] \,,
\label{eq:Liouville_eq_1}
\end{equation}
where $\hat{\tilde{V}}^{\prime}_\mathrm{ext,\mathbf{q}}(\omega)$ is the external perturbing potential, $\hat{\rho}^{\circ}$ is the unperturbed $2 \times 2$ spin-charge density matrix operator, and $\hat{\mathcal{L}}_{\mathbf q}$ is the Liouvillian superoperator, the action of which is defined as:
\begin{equation}
\hat{\mathcal{L}}_\mathbf{q} \cdot \hat{\tilde{\rho}}'_{\mathbf q}(\omega) \equiv
\left[\hat{H}^{\circ}, \hat{\tilde{\rho}}_{\mathbf q}'(\omega)\right] +
\left[\hat{\tilde{V}}^{\prime}_{\mathrm{HXC},\mathbf{q}}[\hat{\tilde{\rho}}_{\mathbf q}'(\omega)], \hat{\rho}^\circ\right] \,,
\label{eq:Liouvillian_def}
\end{equation}
where $\hat{\tilde{V}}^{\prime}_{\mathrm{HXC},\mathbf{q}}$ is the response HXC potential
[see Eq.~\eqref{eq:V_HXC_resp_q}].

The magnetization-density response linearly induced by 
the external magnetic perturbing potential at a specific transferred momentum 
$\mathbf{q}$ and at a specific frequency $\omega$ can be defined as:
\begin{eqnarray}
\bigl\langle \hat{{\boldsymbol m}}_\mathbf{q}' \bigr\rangle_\omega & = &
\mathrm{Tr}[\hat{\boldsymbol m}^\dagger_\mathbf{q} \,
\hat{\tilde{\rho}}_{\mathbf{q}}'(\omega)] \nonumber \\
& = & \left( \hat{\boldsymbol m}_\mathbf{q}, (\omega - \hat{\mathcal{L}}_\mathbf{q})^{-1} \cdot
[\hat{\tilde{V}}^{\prime}_\mathrm{ext, \mathbf{q}}(\omega), \hat{\rho}^\circ] \right) \,,
\label{eq:m_expectation_value}
\end{eqnarray}
where with $(\cdot,\cdot)$ we indicate a scalar product in an operator space.
Using the following convention for the external perturbing potential~\cite{Gorni:2018}
\begin{equation}
\hat{\tilde{V}}^{\prime}_\mathrm{ext,\mathbf{q}}(\omega) =
\hat{{\boldsymbol m}}_\mathbf{q} \cdot \tilde{\boldsymbol b}'_{\mathrm{ext},\mathbf{q}}(\omega),
\label{eq:V_to_m}
\end{equation}
we can rewrite the expectation value~\eqref{eq:m_expectation_value} as
\begin{equation}
\bigl\langle \hat{{\boldsymbol m}}_\mathbf{q}' \bigr\rangle_\omega =
 {\boldsymbol \chi}(\mathbf{q}, \mathbf{q}; \omega) \,
\tilde{\boldsymbol b}'_{\mathrm{ext},\mathbf{q}}(\omega) \,,
\end{equation}
where ${\boldsymbol \chi}(\mathbf{q}, \mathbf{q}; \omega)$ is the $3 \times 3$ spin susceptibility matrix, which reads:
\begin{equation}
{\boldsymbol \chi}(\mathbf{q}, \mathbf{q}; \omega) =
\left( \hat{{\boldsymbol m}}_\mathbf{q}, (\omega - \hat{\mathcal{L}}_\mathbf{q})^{-1} \cdot
[\hat{{\boldsymbol m}}_\mathbf{q}, \hat{\rho}^\circ] \right) \,.
\label{eq:spin-density_susceptibility_2}
\end{equation}
The poles of this quantity mark the magnetic excitations of the system, and they allow to characterize the cross section of numerous magnetic spectroscopies, both bulk ones such as INSS [Eqs.~\eqref{eq:cross_section}--\eqref{eq:S_def}], or surface ones such as SPEELS~\cite{Gokhale:1992}.
Moreover, the usage of a noncollinear framework allows us to take into account the spin-orbit coupling effect as well as to study systems with complex noncollinear patterns in the ground state.
Finally, it is worth noting that our formalism allows us to compute the whole $4 \times 4$ generalized susceptibility matrix which contains spin-spin [Eq.~\eqref{eq:spin-density_susceptibility_2}], charge-charge, spin-charge, and charge-spin couplings (see Appendix~A in Ref.~\cite{Gorni:2018}).
%


\subsection{Batch representation}
\label{sec:batch}

Equations~\eqref{eq:charge_dens_resp_q} and \eqref{eq:spin_dens_resp_q} show that
the response charge and magnetization densities are uniquely determined by the two sets of spinor wave functions $X_{\mathbf{q}} = \{ x_{n,\mathbf{k}+\mathbf{q}} \}$ and
$Y_{\mathbf{q}} = \{ y_{n,\mathbf{k}+\mathbf{q}} \}$,
which are called respectively {\it upper} and {\it lower} components of the {\it batch representation} (BR) of the response spin-charge density matrix operator:
\begin{equation}
\hat{\tilde{\rho}}'_{\mathbf{q}} \xrightarrow{\mathrm{BR}}
\left(
\begin{array}{c}
X_{\mathbf{q}} \\ [5pt]
Y_{\mathbf{q}}
\end{array}
\right) =
\left(
\begin{array}{c}
\{ \tilde{U}'_{n,\mathbf{k}+\mathbf{q}}(\mathbf{r},\omega) \} \\ [5pt]
\{ \hat{\mathrm{T}}\tilde{U}'_{n,-\mathbf{k}-\mathbf{q}}(\mathbf{r},-\omega) \}
\end{array}
\right) \,.
\label{eq:batch_rho}
\end{equation}
This mapping can be formalized by defining BR of a generic operator $\hat{O}_\mathrm{\mathbf{q}}(\omega)$ as
\begin{align}
  \hat{O}_\mathrm{\mathbf{q}}(\omega)
  & \xrightarrow{\mathrm{BR}}\left(
\begin{array}{c}
O_{\mathbf{q}}^X \\[4pt]
O_{\mathbf{q}}^{Y}
\end{array}
\right) \nonumber \\
  & ~~~ = \quad \left(
  \begin{array}{c} \left\{ \hat{P}_{\mathbf{k}+\mathbf{q}} \,
  \hat{O}_\mathrm{\mathbf{q}}(\omega) \,
  U^\circ_{n,\mathbf{k}}(\mathbf{r}) \right\} \\ [6pt]
  \left\{ \hat{\mathrm{T}} \hat{P}_{-\mathbf{k}-\mathbf{q}} \,
  \hat{O}^{\dag}_\mathrm{\mathbf{q}}(\omega) \,
   U^\circ_{n,-\mathbf{k}}(\mathbf{r}) \right\}
\end{array}
\right) \,,
\label{eq:commutator_BR}
\end{align}
similarly to how it is done in Refs.~\cite{Rocca:2008, Malcioglu:2011}. 
Therefore, the commutator appearing on the right-hand side of 
Eq.~\eqref{eq:Liouville_eq_1} in BR will result in:
\begin{align}
  [\hat{\tilde{V}}^{\prime}_\mathrm{ext,\mathbf{q}}, \hat{\rho}^\circ]
  & \xrightarrow{\mathrm{BR}}\left(
\begin{array}{c}
V_{\mathbf{q}}^X \\[4pt]
V_{\mathbf{q}}^{Y}
\end{array}
\right) \nonumber \\
  & ~~~ = \quad \left(
  \begin{array}{c} \left\{ \hat{P}_{\mathbf{k}+\mathbf{q}} \,
  \hat{\tilde{V}}^{\prime}_\mathrm{ext,\mathbf{q}} \,
  U^\circ_{n,\mathbf{k}}(\mathbf{r}) \right\} \\ [6pt]
  \left\{ - \hat{P}^{+}_{\mathbf{k}+\mathbf{q}} \,
  \hat{\tilde{V}}^{\prime +}_\mathrm{ext,\mathbf{q}} \,
  \hat{\mathrm{T}} U^\circ_{n,-\mathbf{k}}(\mathbf{r}) \right\}
\end{array}
\right) \,.
\label{eq:commutator_BR}
\end{align}
Thus, the quantum Liouville equation~\eqref{eq:Liouville_eq_1}
[or equivalently Eqs.~\eqref{eq:KS_eq_resonant_projected_q} and
\eqref{eq:KS_eq_antiresonant_projected_q}] in BR takes the following form:
\begin{equation}
(\omega - \hat{\mathcal{L}_\mathbf{q}})
\left(
\begin{array}{c}
X_{\mathbf{q}} \\
Y_{\mathbf{q}}
\end{array}
\right) =
\left(
\begin{array}{c}
V_{\mathbf{q}}^X \\[4pt]
V_{\mathbf{q}}^{Y}
\end{array}
\right) \,,
\label{eq:Liouville_eq_BR}
\end{equation}
and the Liouvillian in BR reads:
\begin{equation}
\hat{\mathcal{L}}_\mathbf{q} \xrightarrow{\mathrm{BR}}
\left(
\begin{array}{cc}
  \mathcal{D}^{XX}_\mathbf{q} + \mathcal{K}^{XX}_\mathbf{q}  & \mathcal{K}^{XY}_\mathbf{q} \\ [6pt]
 -\mathcal{K}^{YX}_\mathbf{q}     &  -\mathcal{D}^{YY}_\mathbf{q} - \mathcal{K}^{YY}_\mathbf{q}
\end{array}
\right) \,,
\label{eq:Liouvillian_BR_q}
\end{equation}
where the actions of the superoperators, appearing in Eq.~\eqref{eq:Liouvillian_BR_q},
on the response batches are defined as:
  \begin{align}
    \mathcal{D}^{XX}_\mathbf{q} X_\mathbf{q} & \equiv \left\{ (\hat{H}^{\circ}_{\mathbf{k+q}} - \varepsilon^\circ_{n,\mathbf{k}}) \, x_{n,\mathbf{k+q}} \right\} \,,
    \\
    \mathcal{D}^{YY}_\mathbf{q} Y_\mathbf{q} & \equiv \left\{ (\hat{H}^{\circ +}_{\mathbf{k+q}} - \varepsilon^\circ_{n,-\mathbf{k}}) \, y_{n,\mathbf{k+q}} \right\} \,,
    \\
    \mathcal{K}^{XX}_\mathbf{q} X_\mathbf{q} +\mathcal{K}^{XY}_\mathbf{q} Y_\mathbf{q} & \equiv \left\{ \hat{P}_{\mathbf{k+q}} \hat{\tilde{V}}^{\prime}_{\mathrm{HXC},\mathbf{q}}\bigl[\{x_{n,\mathbf{k+q}}\}, \{y_{n,\mathbf{k+q}}\}\bigr] \, U^\circ_{n,\mathbf{k}}(\mathbf{r}) \right\} \,,
    \\
    \mathcal{K}^{YX}_\mathbf{q} X_\mathbf{q} +\mathcal{K}^{YY}_\mathbf{q} Y_\mathbf{q} & \equiv \left\{ \hat{P}^{+}_{\mathbf{k+q}} \hat{\tilde{V}}^{\prime +}_{\mathrm{HXC},\mathbf{q}}\bigl[\{x_{n,\mathbf{k+q}}\}, \{y_{n,\mathbf{k+q}}\}\bigr] \, \hat{\mathrm{T}} U^\circ_{n,-\mathbf{k}}(\mathbf{r}) \right\} \,.
    \label{eq:liouvillian_blocks}
  \end{align}
It is worth noting that due to the lack of time-reversal symmetry,
it is not useful to make a rotation of the batches as it was done for other
spectroscopies~\cite{Rocca:2008, Timrov:2013}.

Finally, we can formally represent $\hat{\boldsymbol m}_\mathbf{q}$ in BR as:
\begin{equation}
\hat{\boldsymbol m}_\mathbf{q} \xrightarrow{\rm BR}
\left(
\begin{array}{c}
{\boldsymbol m}_{\mathbf{q}}^{X} \\ [5pt]
{\boldsymbol m}_{\mathbf{q}}^{Y}
\end{array}
\right) =
\left(
\begin{array}{c}
\{ \hat{P}_{\mathbf{k}+\mathbf{q}} \, \hat{\boldsymbol m}_\mathbf{q} U^\circ_{n,\mathbf{k}}(\mathbf{r}) \} \\ [5pt]
\{ \hat{P}^{+}_{\mathbf{k}+\mathbf{q}} \, \hat{\boldsymbol m}_\mathbf{q} \hat{\mathrm{T}} U^\circ_{n,-\mathbf{k}}(\mathbf{r}) \}
\end{array}
\right) \,.
\label{eq:m_BR}
\end{equation}
Therefore, using Eq.~\eqref{eq:V_to_m} in Eq.~\eqref{eq:commutator_BR}, and using 
Eqs.~\eqref{eq:Liouvillian_BR_q} -- \eqref{eq:m_BR}, we can write the spin
susceptibility matrix~\eqref{eq:spin-density_susceptibility_2} in BR as:
\begin{equation}
{\boldsymbol \chi}(\mathbf{q}, \mathbf{q}; \omega) =
\left( ({\boldsymbol m}_{\mathbf{q}}^{X}, {\boldsymbol m}_{\mathbf{q}}^{Y})^\top, (\omega -
\hat{\mathcal{L}}_\mathbf{q})^{-1} \cdot
({\boldsymbol m}_{\mathbf{q}}^{X}, - {\boldsymbol m}_{\mathbf{q}}^{Y})^\top \right) \,,
\label{eq:spin-density_susceptibility_3}
\end{equation}
which can be efficiently computed using the Lanczos recursion algorithm, as explained in the next section.


\subsection{Lanczos recursion algorithm}
\label{sec:Lanczos_algorithm}

In order to compute the spin susceptibility matrix ${\boldsymbol \chi}(\mathbf{q}, \mathbf{q}; \omega)$ using Eq.~\eqref{eq:spin-density_susceptibility_3}, we need to evaluate the off-diagonal matrix element of the resolvent of the Liouvillian, $(\omega - \hat{\mathcal{L}}_\mathbf{q})^{-1}$. A straightforward inversion of such a matrix in plane-wave framework is a formidable task. It is therefore convenient to use recursive algorithms, such as e.g. the {\it Lanczos recursion algorithm}, which does not rely on the inversion of the matrices, but a recursive evaluation of an off-diagonal matrix element as in Eq.~\eqref{eq:spin-density_susceptibility_3} \cite{Saad:2003}. We will briefly review the two flavors of the Lanczos algorithm that are implemented in the \turboMagnon code, namely, the {\it non-Hermitian Lanczos biorthogonalization algorithm}~\cite{Rocca:2008, Malcioglu:2011, Baroni:2012}, and the {\it pseudo-Hermitian Lanczos algorithm}~\cite{Ge:2014, Gruning:2011, Mostafazadeh:2002}. A more detailed description of the algorithms can be found in the corresponding references. 

According to Eq.~\eqref{eq:spin-density_susceptibility_3}, each element of the spin susceptibility matrix can be written in the form
\begin{equation}
\chi_{\lambda\mu}(\mathbf{q}, \mathbf{q}; \omega)
=
( u^{\lambda} , (\omega - \hat{\mathcal{L}}_{\mathbf{q}})^{-1} v^{\mu}_{\mathbf{q}} )
\, ,
\label{eq:off-diag-resolvent}
\end{equation}
where $\lambda$ and $\mu$ are the indices labeling Cartesian components, and the specific matrix element $\chi_{\lambda\mu}$ depends on the $u^{\lambda}$ and $v^{\mu}_{\mathbf{q}}$ vectors, which in the BR read:
\begin{align}
       u^{\lambda} &= \left( \{ 
 \sigma^{\lambda} U_{n,\mathbf{k}}^\circ(\mathbf{r}) \} ,
 \{  -\sigma^{\lambda} \hat{\mathrm{T}} U_{n,-\mathbf{k}}^\circ(\mathbf{r})   \}  \right)^\top 
 ,
 \\
  v^{\mu}_{\mathbf{q}} &= \left( \{ 
\hat{P}_{\mathbf{k}+\mathbf{q}} \sigma^{\mu} U_{n,\mathbf{k}}^\circ(\mathbf{r}) \} ,
 \{  \hat{P}^{+}_{\mathbf{k}+\mathbf{q}} \sigma^{\mu} \hat{\mathrm{T}} U_{n,-\mathbf{k}}^\circ(\mathbf{r}) \} \right)^\top .
\end{align}
In the non-Hermitian Lanczos biorthogonalization algorithm, by starting from the initial pair of Lanczos vectors $q_1=p_1=v_\mathbf{q}^\mu$ [see Eq.~\eqref{eq:commutator_BR}], two coupled Lanczos chains are generated by recursively applying $\hat{\mathcal{L}}_\mathbf{q}$ and $\hat{\mathcal{L}}^{\dagger}_\mathbf{q}$ to the previous Lanczos chain vectors, $q_i$ and $p_i$~\cite{Rocca:2008, Timrov:2013}.  We note that $q_i$ and $p_i$ implicitly depend on the $\mu$ and $\mathbf{q}$ indices. A pair of biorthogonal basis sets of increasing dimension are thus recursively constructed, $\{ q_i \}$ and $\{ p_i \}$, where $i=\overline{1,M}$, and $M$ being the number of Lanczos iterations. The Lanczos coefficients, $\alpha^\mu_{i,\mathbf{q}}$, $\beta^\mu_{i,\mathbf{q}}$, and $\gamma^\mu_{i,\mathbf{q}}$, are thus computed at each iteration to form a sparse $M$-dimensional tridiagonal matrix, $\Tmatrix$, which is an oblique projection of the Liouvillian onto such biorthogonal bases: $\left( \Tmatrix \right)_{ij} = (p_i, \hat{\mathcal{L}}_\mathbf{q} \, q_j)$.

In order to speed-up the Lanczos recursion, one can take advantage of the pseudo-Hermiticity of the Liouvillian superoperator~\cite{Gruning:2011}.
In this case, by defining a proper metric of the linear space, it is possible to recover a standard Hermitian Lanczos algorithm with a modified scalar product~\cite{Ge:2014, Gruning:2011, Mostafazadeh:2002}, which requires the application of the Liouvillian superoperator only once per Lanczos iteration, resulting in a factor of two speed-up~\cite{Ge:2014}.
As a result of the use of this algorithm one also generates the tridiagonal matrix $\Tmatrix$.

After generating the tridiagonal matrix, $\Tmatrix$, the spin susceptibility matrix~\eqref{eq:off-diag-resolvent} can be computed as~\cite{Timrov:2015}: 
\begin{equation}
  \chi_{\lambda\mu}(\mathbf{q}, \mathbf{q}; \omega) \simeq \left(
  \bm{z}_\mathbf{q}^{\lambda\mu} , ( \omega \mathbb{I} - \Tmatrix)^{-1} \cdot \, \bm{e}_1 \right) , 
  \label{eq:resolvent_Liouvillian_Lanczos_q}
\end{equation}
where $\mathbb{I}$ is the $M$-dimensional unit matrix, $\bm{e}_1 = (1,0,\ldots,0)$ is the $M$-dimensional unit vector, and $\bm{z}_\mathbf{q}^{\lambda\mu} = (z_{1,\mathbf{q}}^{\lambda\mu}, z_{2,\mathbf{q}}^{\lambda\mu}, \ldots, z_{M,\mathbf{q}}^{\lambda\mu})$ is the $M$-dimensional array whose coefficients $z_{i,\mathbf{q}}^{\lambda\mu}$ are computed {\it on-the-fly} of the Lanczos recursion and they are defined as $z_{i,\mathbf{q}}^{\lambda\mu} = (q_i,u^\lambda)$, and the $\mu$ and $\mathbf{q}$ dependence implicitly come from $q_i$. In practice, the right-hand side of Eq.~\eqref{eq:resolvent_Liouvillian_Lanczos_q} is computed by solving the equation:
\begin{equation}
  \left( \omega \mathbb{I} - \Tmatrix \right) \bm{\eta}^\mu_\mathbf{q,\omega} = \bm{e}_1 ,
  \label{eq:post_processing}
\end{equation}
where $\bm{\eta}^\mu_\mathbf{q,\omega}$ is the $M$-dimensional vector which is the solution of the equation above at a fixed value of $\omega$, and finally calculating the scalar product $\chi_{\lambda\mu}(\mathbf{q}, \mathbf{q}; \omega) \simeq  \left( \bm{z}^{\lambda\mu}_\mathbf{q}, \bm{\eta}^\mu_\mathbf{q,\omega} \right )$, which are both inexpensive operations from the computational point of view. This allows one to use very dense low-frequency grids for a very accurate sampling of magnons or extended high-frequency grids for exploring Stoner excitations. The convergence of these spectra with respect to the number of Lanczos iterations, $M$, can be sped up by making use of the extrapolation technique for the Lanczos coefficients, which is described in detail in Refs.~\cite{Rocca:2008, Malcioglu:2011}. More details about the Lanczos algorithm for computing magnetic spectra can be found in Ref.~\cite{Gorni:2016}.
Finally, we note that in practice when solving Eq.~\eqref{eq:post_processing}, a small imaginary part $\eta$ is added to the frequency argument, $\omega \rightarrow \omega + i\eta$, so as to regularize the spin susceptibility matrix $\chi_{\lambda\mu}(\mathbf{q}, \mathbf{q}; \omega+i\eta)$~\cite{Rocca:2008, Baroni:2012, Gorni:2016}.

\subsection{Sum rules}
\label{sec:sum_rules}

As in the case of collective charge excitations~\cite{Timrov:2015}, collective spin excitations also satisfy certain sum rules~\cite{Skovhus:2021}. To show this, we introduce the anti-Hermitian part of the susceptibility tensor
\begin{align}
L_{\lambda\mu}(\mathbf{q},\omega\!+\! i\eta) =
\frac{1}{2 i} \Big[ \chi_{\lambda\mu}(\mathbf{q},\mathbf{q};\omega\!+\! i\eta) -
\chi^*_{\mu\lambda}(\mathbf{q},\mathbf{q};\omega\!+\! i\eta) \Big] \,.
\label{eq:L_def}
\end{align}
It can be shown that $L_{\lambda\mu}(\mathbf{q},\omega\!+\! i\eta)$ satisfies the following sum rule:
\begin{equation}
-\frac{\hbar}{\pi}
\int_{-\infty}^{\infty} L_{\lambda\mu}(\mathbf{q},\omega\!+\! i\eta) d\omega =
2i\mu_{\rm B} \sum_{\nu} \epsilon_{\lambda\mu\nu}
\langle \hat{m}_{\nu}\rangle \,,
\label{eq:sum_rule_1}
\end{equation}
where $\langle \hat{m}_{\nu}\rangle$ is the expectation value of the $\nu$ Cartesian component of the total ground-state magnetization in the unit cell, and $\epsilon_{\lambda\mu\nu}$ is the Levi-Civita tensor. It is important to note that Eq.~\eqref{eq:sum_rule_1} holds for any value of the transferred momentum~$\mathbf{q}$ and for any broadening parameter $\eta$. In addition, we note that the real part of the spin susceptibility matrix is even with respect to $\omega$ while the imaginary part is odd, which implies that only the real part is involved in the sum rule. Finally, by using Eqs.~\eqref{eq:L_def} and \eqref{eq:sum_rule_1}, we present the expression for the sum rule in the case of $\lambda=x$ and $\mu=y$ that is particularly useful for magnets polarized along the $z$ direction, which will be used later in Sec.~\ref{sec:benchmarking} for the CrI$_3$ monolayer:
\begin{equation}
\frac{\hbar}{2\pi\mu_{\rm B}} \int_{0}^{\infty} {\rm Re} \Big[
\chi_{xy}(\mathbf{q},\mathbf{q};\omega\!+\! i\eta) - \chi^*_{yx}(\mathbf{q},\mathbf{q};\omega\!+\! i\eta) \Big] d\omega = \langle \hat{m}_{z}\rangle \,.
\label{eq:sum-rule-reciprocal_space}
\end{equation}
%


\section{Description of software components}
\label{sec:Description_of_software_components}

The \turboMagnon code is designed as a module of the \QE\, distribution~\cite{Giannozzi:2009, Giannozzi:2017, Giannozzi:2020}, and it resides in a self-contained directory \texttt{TDDFPT} under the root directory of the \QE\, tree, which contains also the \texttt{turboTDDFT} and \texttt{turboEELS} codes for the calculation of the absorption and electron energy loss spectra, respectively. The \turboMagnon code uses many of the generic linear response routines contained in the \texttt{LR\_Modules} repository. When the \turboMagnon code is installed (see Sec.~\ref{sec:Installation_instructions}), the \texttt{bin/} directory in the \QE\, root contains links to the executable \texttt{turbo\_magnon.x} (the main program) and \texttt{turbo\_spectrum.x} (a post-processing program). 
The code \texttt{turbo\_magnon.x} performs $\mathbf{q}$-specific Lanczos recursions (up to three; one for each column of the spin susceptibility matrix) to obtain Lanczos coefficients $\alpha_{i,\mathbf{q}}^\mu$, $\beta_{i,\mathbf{q}}^\mu$, $\gamma_{i,\mathbf{q}}^\mu$, and $z_{i,\mathbf{q}}^{\lambda\mu}$ coefficients (see Sec.~\ref{sec:Lanczos_algorithm}), and to construct the tridiagonal representation of the Liouvillian, $\Tmatrix$, while \texttt{turbo\_spectrum.x} uses this matrix to calculate the spin susceptibility matrix, $\chi_{\lambda\mu}(\mathbf{q}, \mathbf{q}; \omega)$, according to Eqs.~\eqref{eq:resolvent_Liouvillian_Lanczos_q} and~\eqref{eq:post_processing}.

\subsection{Ground-state calculation}   

In order to compute the magnetic spectra of a system, a standard spin-polarized ground-state DFT calculation has to be performed first, yielding the KS spinor wave functions $U^\circ_{n,\mathbf{k}}(\mathbf{r})$ and KS energies
$\varepsilon^\circ_{n,\mathbf{k}}$ for all occupied states, which allow us to determine the ground-state charge density, $n^\circ(\mathbf{r})$, and magnetization density, ${\boldsymbol m}^\circ(\mathbf{r})$. The information thus obtained is then used as input for the linear-response calculation with the \turboMagnon code. This ground-state calculation is performed by the \texttt{pw.x} code, which is one of the key components of the \QE\, package. In \ref{sec:Sample_input_files} a sample input file for \texttt{pw.x} is shown for the case of a CrI$_3$ monolayer. After successful completion of the ground-state calculation, the \texttt{pw.x} code writes the ground-state KS wave functions, energies, and charge and magnetization densities to disk, together with all relevant information about the system, like geometry, pseudopotentials, etc. This data is used by the \turboMagnon code which reads all this data at program start. Therefore, it is not necessary to redefine the system under study in the input file of \texttt{turbo\_magnon.x}.

\subsection{TDDFpT calculation}   

The linear-response calculation is done using the \texttt{turbo\_magnon.x} code, which performs the Lanczos recursions (see Sec.~\ref{sec:Lanczos_algorithm}) for a given transferred momentum, $\mathbf{q}$, and for a given direction of the perturbing magnetic field (which is controlled by the input parameter \texttt{ipol}). 
This is by far the most time consuming step of the calculation. In \ref{sec:Sample_input_files} a sample input file for \texttt{turbo\_magnon.x} is shown for the case of a CrI$_3$ monolayer. A list of all
input variables of \texttt{turbo\_magnon.x} is given in Table~\ref{tab:Table_input_turbo_magnon.x} of
\ref{sec:Input_variables}. The integer input variable \texttt{itermax} sets up the number of Lanczos iterations, and so determines the dimension $M$ of the tridiagonal matrix, $\Tmatrix$ (see
Sec.~\ref{sec:Lanczos_algorithm}). In fact, one can check whether the number of iterations is sufficient to achieve an adequately converged spectrum only at the post-processing level (see
Sec.~\ref{sec:post-proc}). It is possible to add more iterations to an existing calculation by restarting the \texttt{turbo\_magnon.x} code, setting the parameter \texttt{restart=.true.} and increasing \texttt{itermax}. The strings defined in the input  variables \texttt{prefix} and \texttt{outdir} identify the system data on disk and must correspond to files created by the \texttt{pw.x} code.

The input variables \texttt{q1}, \texttt{q2}, and \texttt{q3} are the three Cartesian components of the transferred momentum, $\mathbf{q}$, specified in units of $2\pi/a$, where $a$ is the lattice parameter specified in the ground-state calculation by \texttt{pw.x}. 

One can choose which flavor of the Lanczos algorithm to use (see Sec.~\ref{sec:Lanczos_algorithm}). By setting \texttt{pseudo\_hermitian=.true.}, the pseudo-Hermitian Lanczos algorithm will be used, otherwise the non-Hermitian Lanczos biorthogonalization algorithm will be used. It is recommended to use the former, because it is two times faster.

During the execution of the \texttt{turbo\_magnon.x} code, a file named \break \texttt{prefix.beta\_gamma\_z.dat} will be written to the \texttt{outdir} directory.
This file contains the Lanczos coefficients $\alpha_{i,\mathbf{q}}^\mu$, $\beta_{i,\mathbf{q}}^\mu$, $\gamma_{i,\mathbf{q}}^\mu$, and $z_{i,\mathbf{q}}^{\lambda\mu}$ coefficients needed for the post-processing calculation. One can use this information for the analysis of the behavior of these coefficients (see Fig.~\ref{fig:coefficients}).

\subsection{Post-processing spectrum calculation}
\label{sec:post-proc}

Once the tridiagonal matrix, $\Tmatrix$, is constructed from the Lanczos coefficients, one can compute the spin susceptibility matrix according to Eqs.~\eqref{eq:resolvent_Liouvillian_Lanczos_q} and~\eqref{eq:post_processing}. This task is performed by the \texttt{turbo\_spectrum.x} program as a post-processing step, which requires negligible amount of the CPU time with respect to \texttt{turbo\_magnon.x}.

The \texttt{turbo\_spectrum.x} program is used also for the calculation of the absorption spectra computed with \texttt{turboTDDFT} and electron energy loss spectra computed using \texttt{turboEELS}. In order to distinguish the different applications, it is necessary to set \texttt{magnons=.true.} in the input for \texttt{turbo\_spectrum.x}. The labels \texttt{prefix} and \texttt{outdir} identify the system on disk and must correspond to files created by the \texttt{turbo\_magnon.x} code. It is necessary to specify the input parameter \texttt{ipol} which controls which column of the spin susceptibility matrix is computed for a given direction of the magnetic field (the direction of the magnetic field is specified in the input file for \texttt{turbo\_magnon.x}).

In \ref{sec:Sample_input_files} a sample input file for \texttt{turbo\_spectrum.x} is shown for the case of a CrI$_3$ monolayer. A list of input variables for the \texttt{turbo\_spectrum.x} program is given in
Table~\ref{tab:Table_input_turbo_spectrum.x} of \ref{sec:Input_variables}.

As it was mentioned in Sec.~\ref{sec:Lanczos_algorithm}, when solving Eq.~\eqref{eq:post_processing}, a small Lorentzian broadening parameter $\eta$ is added to the frequency in order to regularize the spin susceptibility matrix $\chi_{\lambda\mu}(\mathbf{q}, \mathbf{q}; \omega+i\eta)$~\cite{Rocca:2008, Baroni:2012, Gorni:2016}. 
The magnetic spectra can be computed in any frequency range specified by the keywords \texttt{start} and \texttt{end}, with a step of frequency given by the \texttt{increment} parameter. 

The convergence of the spectrum in the desired frequency range can be checked by varying the number of Lanczos coefficients used. This number is set by the input keywords \texttt{itermax0} and \texttt{itermax}. If no extrapolation of Lanczos coefficients is used (\texttt{extrapolation='no'}), then \texttt{itermax}=\texttt{itermax0}. These variables can take values up to the number of iterations which
have been performed using the \texttt{turbo\_magnon.x} code. For a given number of Lanczos iterations, it is possible to improve the convergence of the computed spectra by extrapolating the coefficients \cite{Malcioglu:2011}. Such an extrapolation can either be bi-constant (\texttt{extrapolation='osc'}) or constant (\texttt{extrapolation='constant'})~\cite{Malcioglu:2011}. In this case, the input variable \texttt{itermax0} indicates the number of exact coefficients to be read from file, while \texttt{itermax} is set to a value which can be chosen arbitrarily large without any significant computational cost. Such an extrapolation procedure amounts to increasing the dimension of the tridiagonal matrix, $\Tmatrix$. It is worth noting though that the extrapolation of Lanczos coefficients for magnetic spectroscopies is most useful to converge the Stoner excitations and less relevant for converging magnons (when they are not overlapping with the former).

Finally, the \texttt{turbo\_spectrum.x} program generates a file called \texttt{prefix.plot\_chi.dat} which contains real and imaginary parts of one column (controlled by \texttt{ipol}) of the spin susceptibility matrix, $\chi_{\lambda\mu}(\mathbf{q}, \mathbf{q}; \omega)$, for a given $\mathbf{q}$ and for each value of the frequency $\omega$. This can be directly used to plot the magnetic spectra using Eq.~\eqref{eq:S_def}.


\section{Installation instructions and parallelization of the code}
\label{sec:installation_and_parallelization}

\subsection{Installation instructions}
\label{sec:Installation_instructions}

The \turboMagnon program is distributed as source code, like the other components of the \QE\, distribution. The installation procedure is the same for all modules in the \QE\,
package. \QE\, and \turboMagnon make use of GNU autoconf~\cite{GNUautoconf}. The \texttt{TDDFPT} repository, which contains the source \turboMagnon code 
must be residing within the \QE\, tree. The code is compiled with the following commands from within the \QE\,tree:
\begin{equation*}
 \begin{array}{c}
   \texttt{./configure} \\ 
   \texttt{make pw} \qquad\quad \\
   \texttt{make tddfpt}
 \end{array}
\end{equation*}
Alternatively, it is possible to use \texttt{cmake}~\cite{cmake} instead of \texttt{./configure}.
Here, the first step sets up the environment (compilers, libraries, etc.). The second step compiles the \texttt{pw.x} code and creates a link to this executable in the \texttt{bin/} repository of the \QE\,
tree. In the third step, the \turboMagnon code (\texttt{turbo\_magnon.x} and \texttt{turbo\_spectrum.x}) are compiled, together with the \texttt{turboTDDFT} and \texttt{turboEELS} codes. Links to these programs are created in the \texttt{bin/} directory of the \QE\, tree. Further detailed installation instructions can be found in the documentation that comes with the \QE\, distribution.

\subsection{Parallelization}

Like the other components of the \QE\, package, the \turboMagnon code is optimized to run on massively parallel architectures. The parallelization of the \turboMagnon code is achieved by using the message-passing paradigm and calls to standard Message Passing Interface (MPI) libraries~\cite{MPI:1994}. High performance on massively parallel architectures is achieved by distributing both data and computations in a hierarchical way across processors. The \turboMagnon code supports two levels of parallelization: $\emph{i)}$ a plane-wave parallelization, which is implemented by distributing real- and reciprocal-space grids across the processors, and $\emph{ii)}$~a~$\mathbf{k}$~points parallelization, which is implemented by dividing
all processors into pools, each taking care of one or more $\mathbf{k}$ points. The Fast Fourier Transforms (FFT's), which are used for transformations from real space to reciprocal space and vice versa, are also efficiently parallelized among processors.

%
\section{Benchmarking}
\label{sec:benchmarking}
%

We now proceed to the validation of the \turboMagnon code by calculating the magnetic spectra for a CrI$_3$ monolayer, a 2D ferromagnetic insulator with honeycomb-arranged magnetic moments and strong spin-orbit coupling. We remark that the correctness of the LL approach in the case of collinear metallic systems (bulk Fe and Ni) has already been inspected and validated in our previous publications~\cite{Gorni:2016, Gorni:2018}, and that the goal here is to validate the noncollinear implementation including spin-orbit coupling. Nonetheless, at the end of this section we will inspect the fulfillment of the sum rule not only for CrI$_3$ monolayer but also for bulk Fe and Ni, since this was not discussed before.

\subsection{Technical details}
\label{sec:technical_details}

All the calculations were performed using the \QE\ distribution~\cite{Giannozzi:2009, Giannozzi:2017, Giannozzi:2020}, a suite of computer codes based on plane waves and pseudopotentials. We performed the ground-state noncollinear spin-polarized calculations using the \texttt{pw.x} program; local spin density approximation (LSDA) was used for the XC functional, and SOC was included self-consistently. Fully-relativistic norm-conserving (NC) pseudopotentials (PPs) were taken from the pseudopotential library of Ref.~\cite{THEOSlib} (\texttt{Cr.rel-pz-n-nc.UPF} and \texttt{I.rel-pz-n-nc.UPF}). 
The KS wave functions and potentials were expanded in plane waves up to a kinetic-energy cutoff of 60 and 240~Ry, respectively. The BZ was sampled using a uniform $\Gamma$-centered $8\times 8\times 1$ $\mathbf{k}$ points mesh. 

\begin{figure}[h!]
    \centering
    \includegraphics[width=0.55\textwidth]{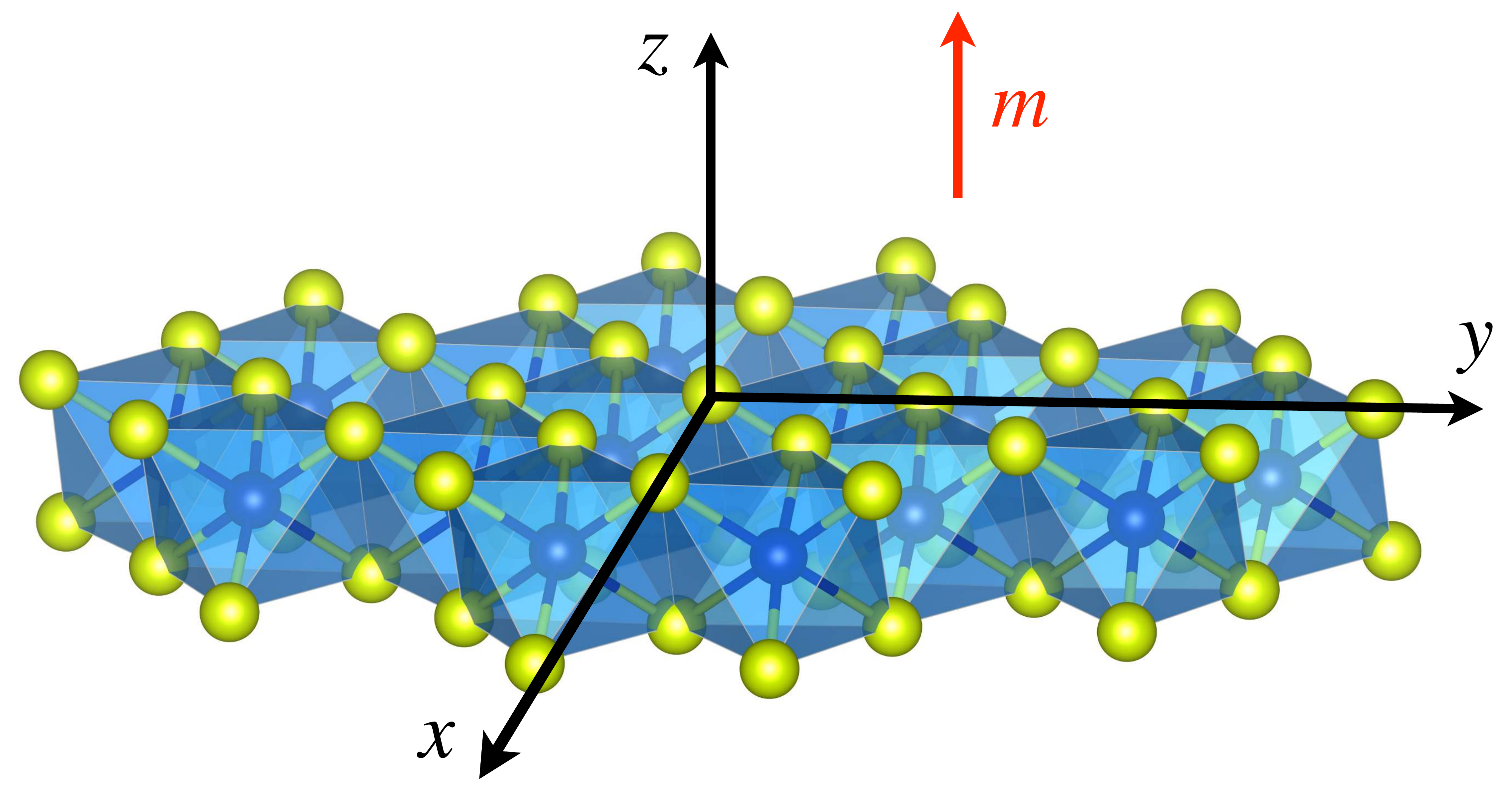}
    \caption{Crystal structure of a CrI$_3$ monolayer. $m = \langle \hat{m}_z \rangle$ is the ground-state magnetization. Cr and I atoms are indicated with blue and yellow balls, respectively.}
    \label{fig:structure}
\end{figure}

The CrI$_3$ 2D crystal structure (see Fig.~\ref{fig:structure}) is described using a honeycomb lattice lying in the $xy$ plane and leaving 17.5~\AA\ of vacuum along the $z$ direction. The monolayer was obtained by extracting one layer from the trigonal bulk structure and by optimizing atomic positions and the in-plane lattice constant (12.98~Bohr). The DFT calculations correctly yield a ferromagnetic (FM) ordering and a magnetic anisotropy with the out-of-plane directions as the easy axis~\cite{Soriano:2020}. The ground-state magnetization is polarized along the $z$ direction as depicted in Fig.~\ref{fig:structure}, and the total magnetic moment is 6~$\mu_{\rm B}$ per unit cell coming mainly from the Cr atoms, consistently with a $S=3/2$ spin on each Cr site. We also quantified the magnetic anisotropy energy: the energy difference between the in-plane and out-of-plane magnetizations is $0.74$~meV. 

TDDFpT calculations were performed using the \texttt{turboMagnon} code (\texttt{turbo\_magnon.x}) in the noncollinear framework consistently with the ground state. Magnetic spectra were obtained using the post-processing program \texttt{turbo\_spectrum.x} using a Lorentzian smearing function with broadening parameters reported in the next sections. The kinetic-energy cutoff and the $\mathbf{k}$ points mesh were used the same as for the ground-state calculation. The convergence of spectra with respect to the number of Lanczos iterations will be shown in the following.

The data used to produce the results of this work are available in the Materials Cloud Archive~\cite{MaterialsCloudArchive2022}.

\subsection{Spin susceptibility matrix}
\label{sec:spin_susceptibility_matrix}

\begin{figure}[h!]
    \centering
    \includegraphics[width=0.49\textwidth,trim= 50 20 35 20,clip]{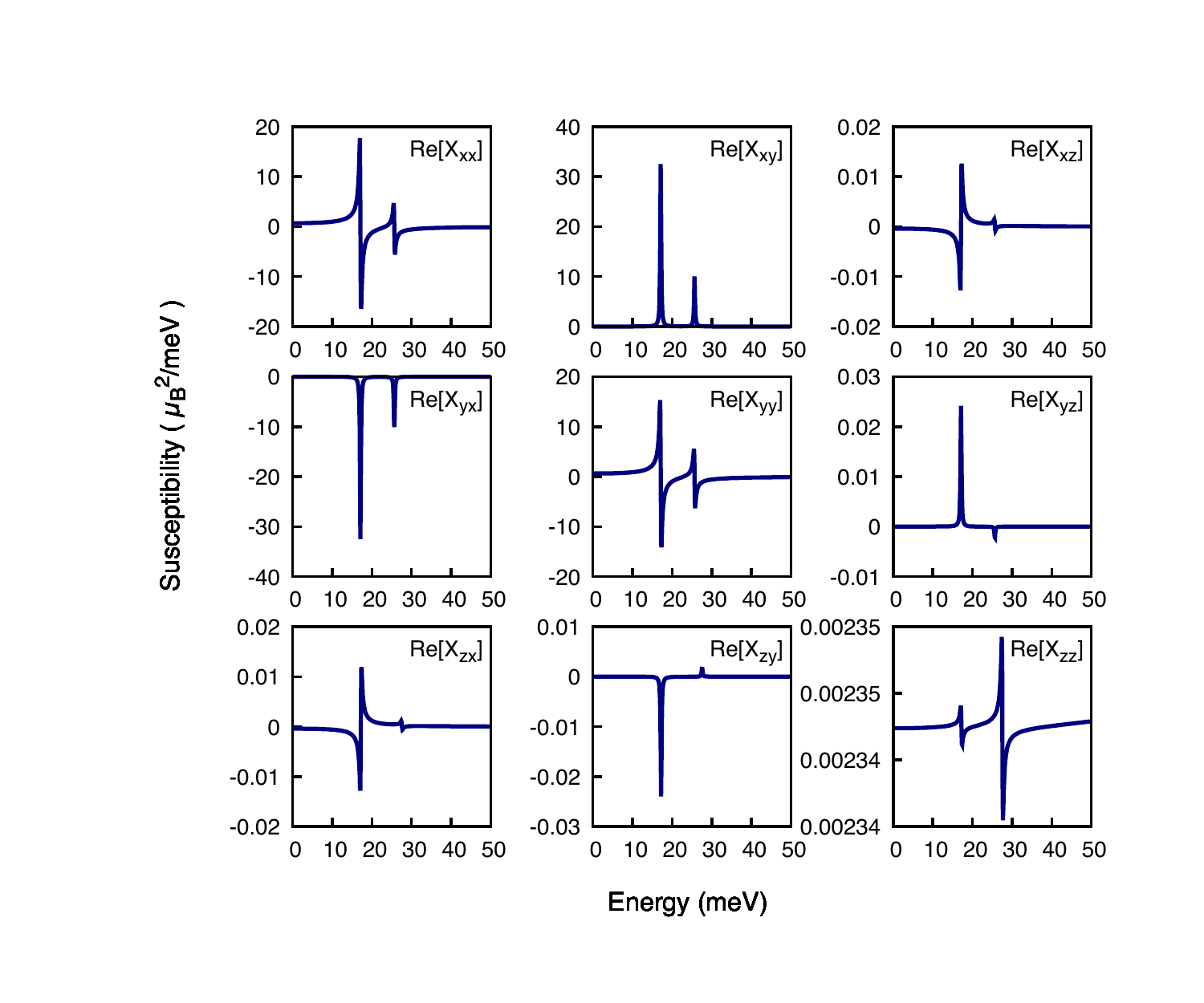}
     \includegraphics[width=0.49\textwidth,trim= 50 20 35 20,clip]{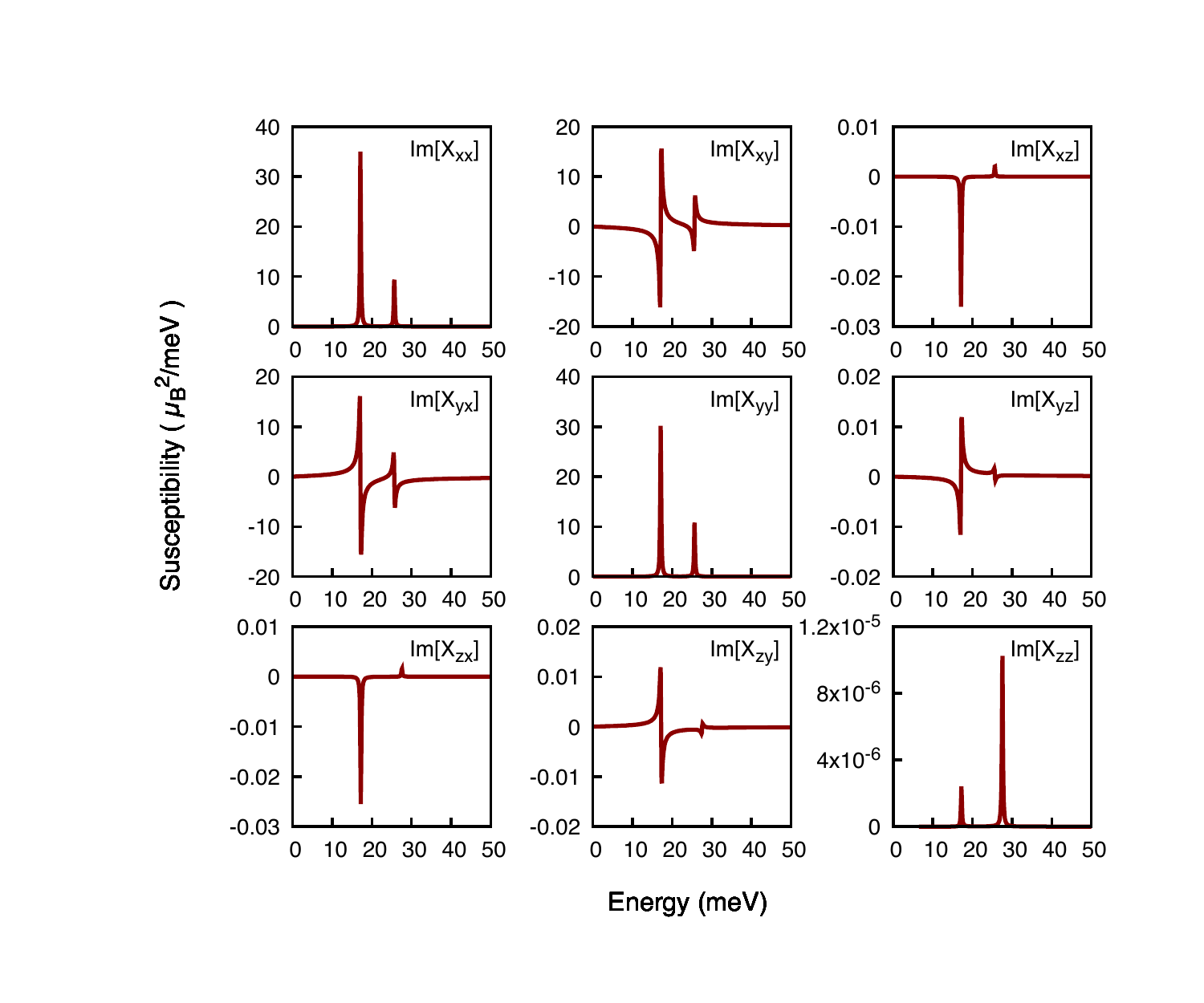}
    \caption{Real (left $3 \times 3$ panel) and imaginary (right $3 \times 3$ panel) parts of the spin susceptibility matrix $\chi_{\lambda\mu}(\mathbf{q}, \mathbf{q}; \omega)$ for a CrI$_3$ monolayer computed using the LL approach to TDDFpT including SOC. The non-Hermitian Lanczos algorithm was used with the Lorentzian broadening $\eta = 1.0$~meV. The transferred momentum $\mathbf{q}$ is equal to the M high-symmetry point in the BZ.}
    \label{fig:3x3susceptibility}
\end{figure}

We start by analyzing the real and imaginary components of the $3 \times 3$ spin susceptibility matrix $\chi_{\lambda\mu}(\mathbf{q},\mathbf{q};\omega)$ computed for the transferred momentum $\mathbf{q}$ being equal to the high-symmetry $M$ point in the BZ. The result for the CrI$_3$ monolayer is shown in Fig.~\ref{fig:3x3susceptibility}. We used the non-Hermitian Lanczos algorithm that was already introduced in our previous work~\cite{Gorni:2018}, but here it is generalized to the noncollinear framework including SOC (this will be discussed in more detail in the following). The goal of this subsection is to analyze and compare all components of the spin susceptibility matrix, and to understand their meaning. We are mainly interested in the imaginary part of $\chi_{\lambda\mu}(\mathbf{q},\mathbf{q};\omega)$ since it is used in Eqs.~\eqref{eq:cross_section} and \eqref{eq:S_def}, and can be directly compared with the double-differential cross sections that are determined from the INSS experiments. But it is also useful to analyze the real parts of $\chi_{\lambda\mu}(\mathbf{q},\mathbf{q};\omega)$ that are used e.g. to check the sum rules (see Sec.~\ref{sec:sum_rules}). 

As can be seen in Fig.~\ref{fig:3x3susceptibility}, the intensities of $\mathrm{Im}[\chi_{xx}]$ and $\mathrm{Im}[\chi_{yy}]$ are 6 orders of magnitude larger than that of $\mathrm{Im}[\chi_{zz}]$. This is so because for the specific geometry setup shown in Fig.~\ref{fig:structure}, $\mathrm{Im}[\chi_{xx}]$ and $\mathrm{Im}[\chi_{yy}]$ are the transversal components of the spin susceptibility matrix that correspond to the excitations of magnons, while $\mathrm{Im}[\chi_{zz}]$ 
describes longitudinal spin excitations, which are considerably stiffer than the former. Despite the huge difference in the intensities, the peak positions in the transversal and longitudinal components of the spin susceptibility appear at similar energies. More specifically, the first peak is at exactly the same energy for all three components (17~meV), while the second peak appears at 26~meV for the two transversal components and at 29~meV for the longitudinal component. Moreover, it is interesting to observe that the first peak is more intense than the second peak both in $\mathrm{Im}[\chi_{xx}]$ and $\mathrm{Im}[\chi_{yy}]$, while the trend is the opposite for $\mathrm{Im}[\chi_{zz}]$. These two peaks in the transversal components of the spin susceptibility matrix correspond to the acoustic and optical magnon excitations, and it appears that the intensity of the acoustic mode is larger than that of the optical mode. It is instructive to analyze also the off-diagonal components of the spin susceptibility matrix. We can see from Fig.~\ref{fig:3x3susceptibility} that these show symmetry modulo the sign. $\mathrm{Im}[\chi_{xy}]$ and $\mathrm{Im}[\chi_{yx}]$ are 2 times less intense than the diagonal components $\mathrm{Im}[\chi_{xx}]$ and $\mathrm{Im}[\chi_{yy}]$, while $\mathrm{Im}[\chi_{xz}]$, $\mathrm{Im}[\chi_{yz}]$, $\mathrm{Im}[\chi_{zx}]$, and $\mathrm{Im}[\chi_{zy}]$ are $3-4$ orders of magnitude less intense than $\mathrm{Im}[\chi_{xx}]$ and $\mathrm{Im}[\chi_{yy}]$. It is important to note that these latter components, which couple the transversal and longitudinal excitations, are identically zero in collinear magnets in the absence of spin-orbit coupling. Similar trends are seen for the real part of the spin susceptibility matrix in Fig.~\ref{fig:3x3susceptibility}.

From the practical point of view, in order to obtain the full $3 \times 3$ spin susceptibility matrix as shown in Fig.~\ref{fig:3x3susceptibility} one needs to perform 3 Lanczos chains. More specifically, the first column in the $\chi_{\lambda\mu}(\mathbf{q},\mathbf{q};\omega)$ matrix is obtained by performing one Lanczos chain for the external magnetic field applied along the $x$ axis, i.e. $\mu = x$ (\texttt{ipol=1}) and by measuring the magnetic response along the 3 Cartesian directions, i.e. $\lambda=x,y,z$ (this is done automatically by the \turboMagnon code). The second and the third columns of $\chi_{\lambda\mu}(\mathbf{q},\mathbf{q};\omega)$ are obtained by applying the external magnetic field along the $y$ (\texttt{ipol=2}) and $z$ (\texttt{ipol=3}) axes, respectively. 
However, in the case of the setup shown in Fig.~\ref{fig:structure} there is actually no need to perform 3 Lanczos chains but just one either with \texttt{ipol=1} or \texttt{ipol=2}, since we are interested only in the transversal components of the spin susceptibility matrix, and $\mathrm{Im}[\chi_{xx}] = \mathrm{Im}[\chi_{yy}]$. This is very convenient from the computational point of view since we can save the CPU time by a factor of 3.

\subsection{Non-Hermitian vs pseudo-Hermitian Lanczos algorithms} 
\label{sec:pseudo_vs_nonpseudo}

\begin{figure}[h!]
    \centering
    \includegraphics[width=0.6\textwidth]{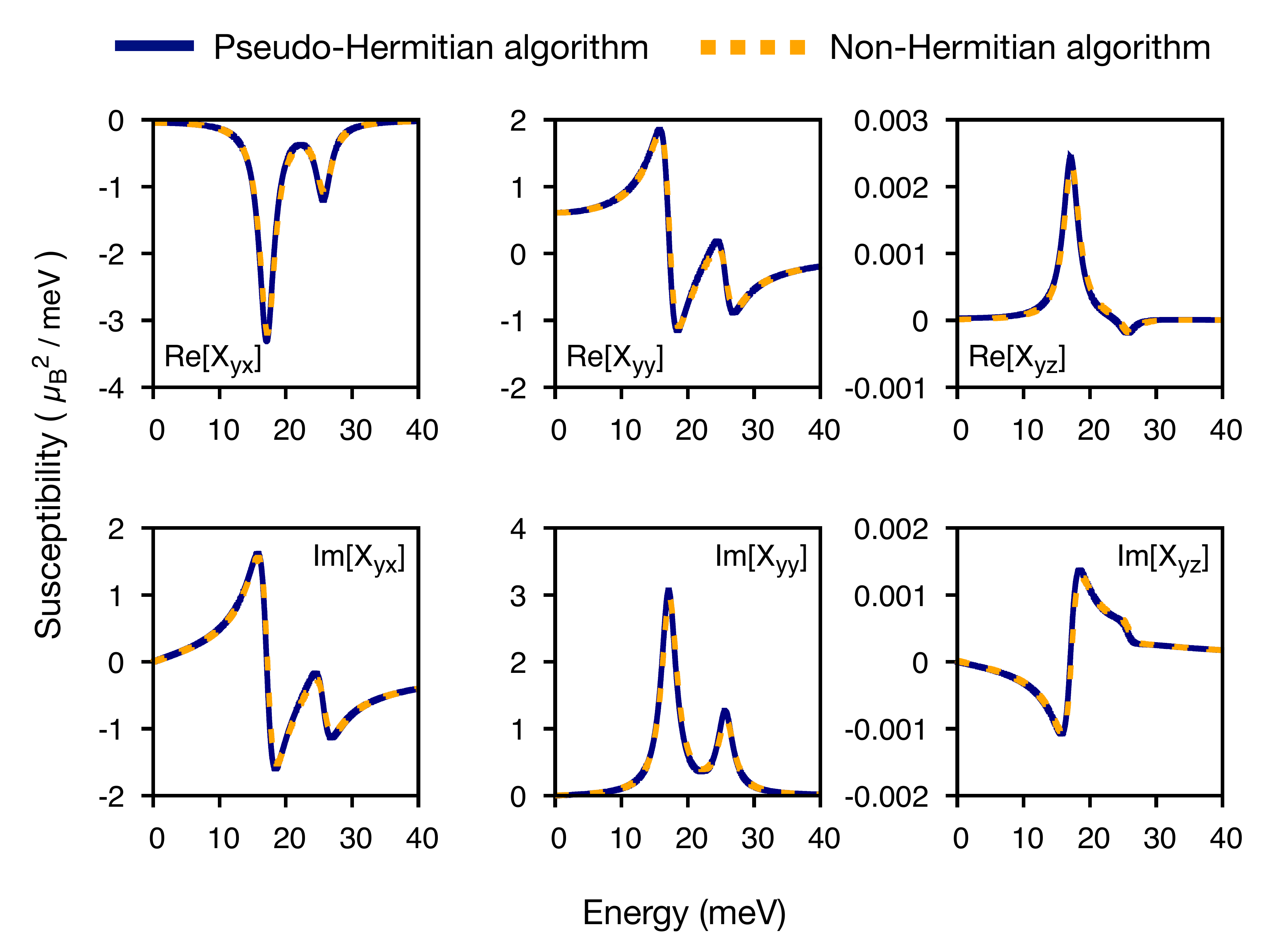}
    \caption{Comparison of the pseudo-Hermitian (solid blue line) and non-Hermitian (dashed orange line) Lanczos algorithms for computing real and imaginary parts of the one row ($\lambda=y$) in the spin susceptibility matrix $\chi_{\lambda\mu}(\mathbf{q}, \mathbf{q}; \omega)$ for a CrI$_3$ monolayer using the LL approach to TDDFpT including SOC. The Lorentzian broadening with $\eta = 1.36$~meV was used. The transferred momentum $\mathbf{q}$ is equal to the M high-symmetry point in the BZ.}
    \label{fig:pseudohermitian}
\end{figure}

We now to proceed to a validation of the new pseudo-Hermitian Lanczos algorithm that was implemented in the \turboMagnon code. As was briefly explained in Sec.~\ref{sec:Lanczos_algorithm}, the pseudo-Hermitian Lanczos algorithm is two times faster than the non-Hermitian one due to the reduction in the number of linear-algebra operations by taking advantage from the pseudo-Hermiticity of the linear-response equations~\cite{Gruning:2011}. In order to check the implementation, in Fig.~\ref{fig:pseudohermitian} we compare the real and imaginary parts of one row ($\lambda=y$) of the spin susceptibility matrix $\chi_{\lambda\mu}(\mathbf{q}, \mathbf{q}; \omega)$ computed using the pseudo-Hermitian and non-Hermitian Lanczos algorithms. For the sake of clearer comparison, we slightly increased the value of the Lorentzian broadening parameter compared to Fig.~\ref{fig:3x3susceptibility}. As can be seen in Fig.~\ref{fig:pseudohermitian}, the two flavors of the Lanczos algorithm give spectra that are in remarkable agreement with each other which validates the correctness of the implementation of the pseudo-Hermitian algorithm. 

It is worth mentioning that both algorithms give converged spectra (with a precision of a fraction of meV) for $\mathbf{q}=\mathrm{M}$ after performing 30000 Lanczos iterations. This is a rather large number of iterations, and hence the numerical stability of the Lanczos chain becomes very relevant (i.e. that Lanczos coefficients do not diverge). In this respect, the pseudo-Hermitian algorithm turns out to be more numerically stable than the non-Hermitian one, therefore the former one should be used by default. Additionally, we want to stress that converging magnetic spectra at different $\mathbf{q}$ requires different number of Lanczos iterations, and as will be shown in Sec.~\ref{sec:convergence} for smaller values of $\mathbf{q}$ we need a smaller number of Lanczos iterations. In the rest of this paper we present results obtained using only the pseudo-Hermitian algorithm.

\subsection{Magnon dispersion and the importance of spin-orbit coupling}
\label{sec:dispersion}

The effect of SOC on magnetic excitations within TDDFT was addressed only in Ref.~\cite{dosSantosDias:2015} to the best of our knowledge. 
Such a scarce availability of TDDFT studies with SOC is partly due to its higher computational cost and partly due to the difficulty in enforcing the correct long-wavelength limit, as it will be explained in the following.
As shown in Sec.~\ref{sec:theory}, our formulation of TDDFpT equations naturally incorporates SOC, allowing us to investigate this rather unexplored subject.
%

\begin{figure}[h!]
    \centering
    \includegraphics[width=0.7\textwidth]{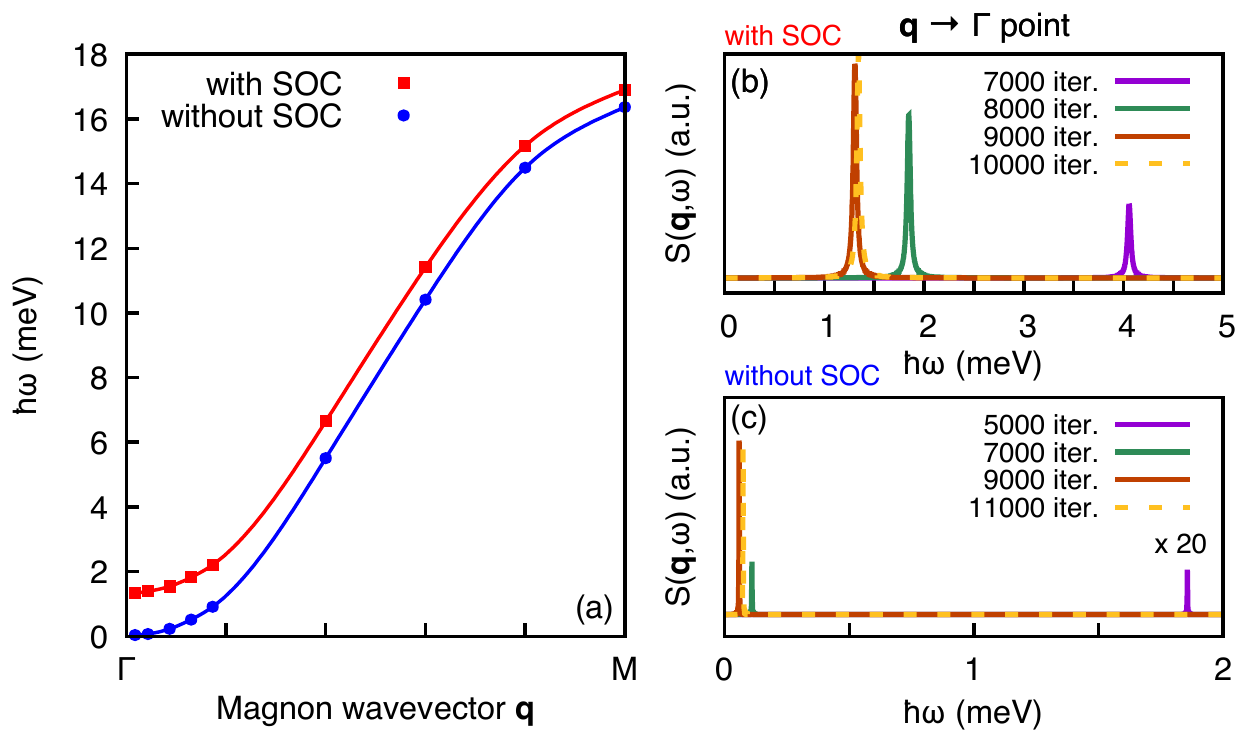}
    \caption{(a)~Magnon energy ($\hbar\omega$) for the CrI$_3$ monolayer along the $\Gamma$-M high-symmetry direction in the BZ with (red squares) and without (blue circles) SOC. The red and blue continuous lines are guides for the eye and they were obtained using spline interpolation. Panels (b) and (c) show the convergence of $S(\mathbf{q},\omega)$ as a function of the number of Lanczos iterations for $|\mathbf{q}| = 0.010 \, (2\pi/a)$ with SOC and $|\mathbf{q}| = 0.025 \, (2\pi/a)$ without SOC along the $\Gamma$-M direction, where $a$ is the lattice parameter. All calculations were performed using the pseudo-Hermitian Lanczos algorithm with a Lorentzian broadening of $\eta = 0.020$~meV in the former case (with SOC) and $\eta = 0.001$~meV in the latter (without SOC).}
    \label{fig:magn_dispersion_and_conv}
\end{figure}

Figure~\ref{fig:magn_dispersion_and_conv}~(a) shows the magnon dispersion for the CrI$_3$ monolayer along the $\Gamma$-M high-symmetry direction in the BZ computed using the pseudo-Hermitian Lanczos algorithm with and without SOC. 
This magnon dispersion represents the $\mathbf{q}$ dependence of the acoustic magnon branch, while the optical magnon branch has vanishing intensity and hence it is not shown on the plot (the optical branch starts having nonvanishing intensities only for $\mathbf{q}$'s approaching the M point). 
The magnon energy at $\mathbf{q} \rightarrow \Gamma$ deserves a special attention due to the Goldstone theorem, which implies that the acoustic magnon energy must be exactly zero in absence of SOC~\cite{Goldstone:1962,Watanabe:2012}.
In practice, the Goldstone theorem is often  violated in actual calculations due to different numerical approximations that are used to describe the ground state and the excited states (e.g. different $\mathbf{k}$ point grids, basis sets, etc.)~\cite{Lounis:2011, Buczek:2011b, Rousseau:2012, Muller:2016}.
Not incurring in the aforementioned limitations, our implementation is expected to satisfy the Goldstone theorem with high accuracy and yield the correct long-wavelength limit. We note that currently the \turboMagnon code does not contain the implementation of the exact limit $\mathbf{q} = \mathbf{0}$ ($\mathbf{q} = \Gamma$), hence in practice we use very small but finite values of $\mathbf{q}$ close to the $\Gamma$ point.
At $\mathbf{q} \to \Gamma$ without SOC we find indeed a vanishing magnon energy, with a value of $0.04$~meV for $|\mathbf{q}| = 0.01 \, (2\pi/a)$, where $a$ is the lattice parameter reported in Sec.~\ref{sec:technical_details}. 
A quadratic fit of the five smallest $|\mathbf{q}|$ points close to $\Gamma$ extrapolates to $\approx 0.02$~meV at $\mathbf{q} = \Gamma$.
This value is orders of magnitude smaller than typical values reported in other works when using other methods (often as large as several tenths of meV~\cite{Muller:2016,Skovhus:2021}).

%

In the presence of SOC, the Goldstone theorem is not supposed to hold since spin-rotational symmetry is already broken, and the acoustic magnon can acquire a finite energy at $\mathbf{q} = \Gamma$, so forming the so-called ``Goldstone gap''. When accounting for spin-orbit coupling, our calculations yield a Goldstone gap of 1.3~meV at $\mathbf{q} \rightarrow \Gamma$ in the CrI$_3$ monolayer.
We note that this value is consistent with a localized spin model where magnetic anisotropies are only onsite: in this case the energy difference between the in-plane and out-of-plane magnetization is estimated to be $0.74$~meV from our ground-state DFT calculations, and the Goldstone gap is twice of that value (i.e. $1.48$~meV)~\cite{Soriano:2020}.
Finally, we notice that the inclusion of SOC does not induce a simple rigid upward shift of the acoustic magnon branch, but has a more complex $\mathbf{q}$-dependent effect, which in a localized spin model implies some renormalization of the intersite exchange coefficients.

No experimental measurements of magnon energies in the CrI$_3$ monolayer have been performed so far. Therefore, we resort to the available INSS measurements in bulk samples in order to give some insights about the predictive accuracy of our calculations.
The computed magnon dispersion along the $\Gamma$--M direction in Fig.~\ref{fig:magn_dispersion_and_conv}~(a) presents the same features and trends as the one for the bulk CrI$_3$ in experiments. We find, however, an overall overestimation of the magnon bandwidth ($\approx 26$\,meV~\cite{Delugas:2021} vs \ $\approx 17$\,meV~\cite{Chen:2018}), which can be related to the known overestimation of magnon stiffness when using ALSDA.
Likewise, the theoretical Goldstone gap in the bulk CrI$_3$ is about $1.3$~meV (which turns out to be the same as in the CrI$_3$ monolayer) and it overestimates the experimental one, whose estimates in CrI$_3$ bulk samples point towards values in the order of 0.37~meV~\cite{Chen:2020}.
Ultimately, to increase the accuracy of quantitative predictions using the LL approach to TDDFpT more advanced XC functionals must be used, which will be the topic of future developments for the \turboMagnon code.


\subsection{Convergence of magnetic spectra and the behavior of Lanczos coefficients}
\label{sec:convergence}

Figures~\ref{fig:magn_dispersion_and_conv}~(b) and (c) show the convergence of $S(\mathbf{q},\omega)$ with and without SOC, respectively. 
We remind that $S(\mathbf{q},\omega)$ is computed using Eq.~\eqref{eq:S_def} which requires on input the spin susceptibility matrix $\chi_{\lambda\mu}(\mathbf{q},\mathbf{q};\omega)$  discussed previously. 
We note that $S(\mathbf{q},\omega)$ takes into account both the transversal and longitudinal magnetic excitations, that are both probed in INSS experiments. In the CrI$_3$ case, however, transverse excitations dominate the response spectrum, so that $S(\mathbf{q},\omega)$ is predominantly a fingerprint of magnon excitations.
We can see in Figs.~\ref{fig:magn_dispersion_and_conv}~(b) and (c) that $S(\mathbf{q},\omega)$ is converged after 10000 Lanczos iterations in both cases. 
We notice, however, that in the case when SOC is included it takes 3 times less Lanczos iterations to convergence the spectrum at $\mathbf{q} \rightarrow \Gamma$ than at $\mathbf{q} = \mathrm{M}$ (see Sec.~\ref{sec:pseudo_vs_nonpseudo}), which means that the convergence is strongly $\mathbf{q}$-dependent. 

\begin{figure}[h!]
    \centering
    \includegraphics[width=0.85\textwidth,trim= 0 80 0 120,clip]{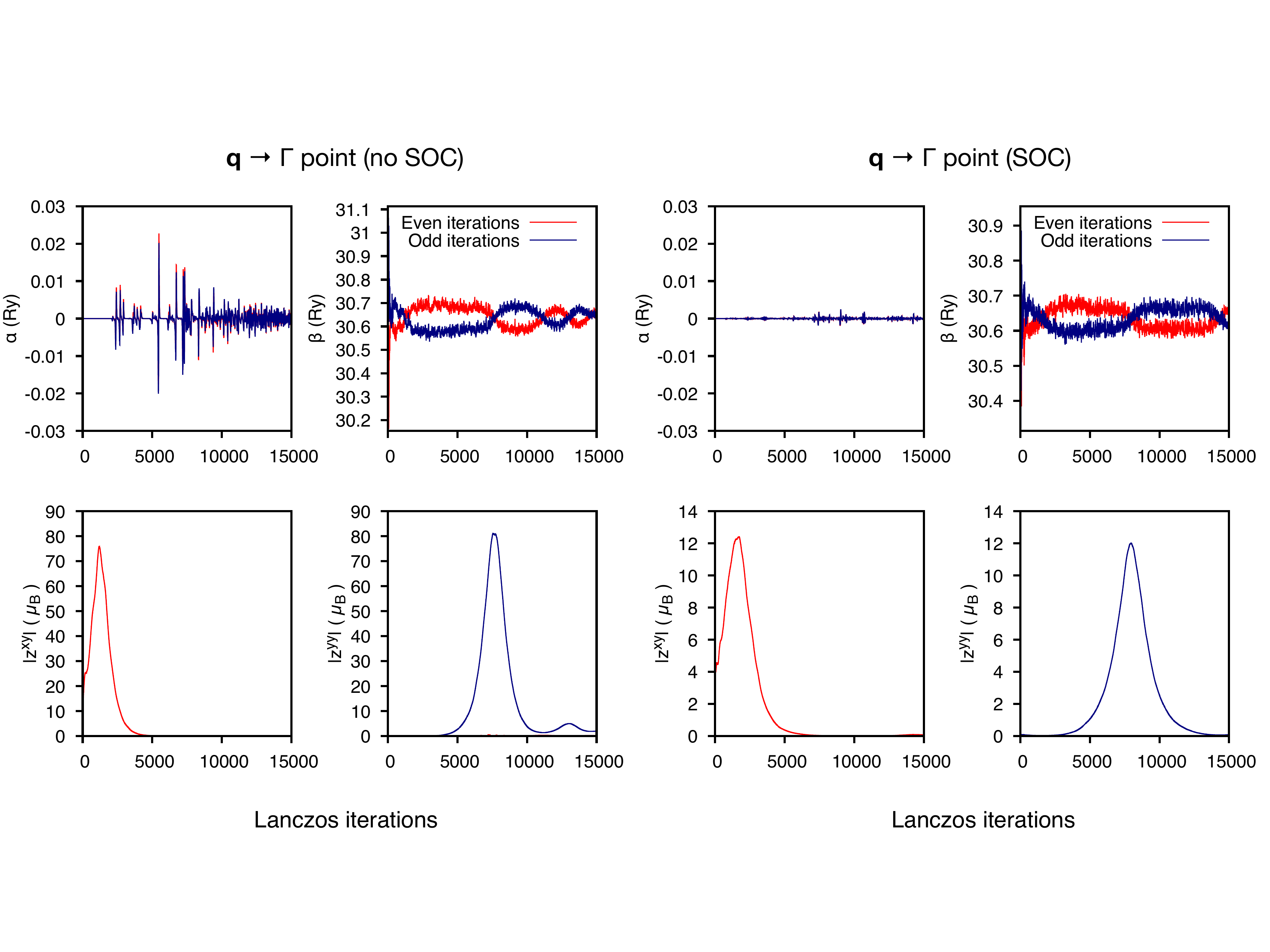}
    \caption{The behavior of Lanczos coefficients $\alpha^\mu_{i,\mathbf{q}}$ and $\beta^\mu_{i,\mathbf{q}}$ (in Ry) and $z^{\lambda\mu}_{i,\mathbf{q}}$ coefficients (in $\mu_\mathrm{B}$) for the CrI$_3$ monolayer as a function of the number of Lanczos iterations $i$. The data is generated using the pseudo-Hermitian Lanczos algorithm at $\mathbf{q} \rightarrow \Gamma$ without SOC (left 4 panels) and with SOC (right 4 panels). The coefficients plotted here correspond to the polarization of the external magnetic field along the $y$ axis (i.e. $\mu=y$); moreover, we note that $z^{zy}$ is vanishing compared to $z^{xy}$ and $z^{yy}$, thus only the latter two are shown. For the sake of simplicity, we dropped certain indices from the coefficients on the plots.}
    \label{fig:coefficients}
\end{figure}

It is also useful to analyze the behavior of the Lanczos coefficients. Figure~\ref{fig:coefficients} shows the behavior of the Lanczos coefficients $\alpha^\mu_{i,\mathbf{q}}$ and $\beta^\mu_{i,\mathbf{q}}$, as well as of the $z^{\lambda\mu}_{i,\mathbf{q}}$ coefficients as a function of the number of Lanczos iterations $i$ using the pseudo-Hermitian algorithm at $\mathbf{q} \rightarrow \Gamma$ with and without SOC when computing the response to a magnetic field polarized along the $y$ axis. The Lanczos coefficients $\gamma^\mu_{i,\mathbf{q}}$ are equal by modulus to $\beta^\mu_{i,\mathbf{q}}$ and can differ only by sign~\cite{Malcioglu:2011}, thus we report only the latter.
First of all, we observe that the $\alpha^\mu_{i,\mathbf{q}}$ coefficients are extremely small compared to the $\beta^\mu_{i,\mathbf{q}}$ coefficients, which was also found in the case of bulk Fe and Ni~\cite{Gorni:2018}.
In fact, the $\alpha^\mu_{i,\mathbf{q}}$ coefficients are exactly zero in the case of the absorption~\cite{Rocca:2008, Malcioglu:2011} and electron energy loss spectroscopies~\cite{Timrov:2013, Timrov:2015}, due to the possibility to perform a rotation to the standard batch representation. Such an argument no longer holds when magnetic ground states are considered and, even though in practice we always find $\alpha^\mu_{i,\mathbf{q}}$ to be very small, we chose not to constrain their value in the absence of a formal proof.
As for the $\beta^\mu_{i,\mathbf{q}}$ coefficients, we show even and odd values as was also done in previous works~\cite{Rocca:2008, Malcioglu:2011, Timrov:2015, Gorni:2018}. We find that the even and odd $\beta$'s oscillate around a mean value that is approximately equal to half the value of the kinetic-energy cutoff ($\sim 60/2 = 30$~Ry). As we discussed in Sec.~\ref{sec:dispersion}, in Figs.~\ref{fig:magn_dispersion_and_conv}~(b) and (c) we see that the magnon peaks are converged after about 10000 Lanczos iterations. From the respective panels in Fig.~\ref{fig:coefficients} we see that the Lanczos $\beta^\mu_{i,\mathbf{q}}$ coefficients have oscillatory behavior and that after 10000 Lanczos iterations signatures of stabilization are observed. 

Finally, we analyze the behavior of the $z^{\lambda\mu}_{i,\mathbf{q}}$ coefficients, which appear to be peaked functions as can be seen in Fig.~\ref{fig:coefficients} (lower panels). Interestingly, the $z^{xy}_{i,\mathbf{q}}$ components are peaked much earlier than $z^{yy}_{i,\mathbf{q}}$ with respect to the number of Lanczos iterations. The $z^{yy}_{i,\mathbf{q}}$ coefficients are peaked at $\sim 8000$ Lanczos iterations both without and with SOC, while the full convergence of the magnon peaks is reached a couple of thousands of Lanczos iterations later. This means that in order to convergence the magnon energies it is necessary not only to reach the point when the $\beta^\mu_{i,\mathbf{q}}$ Lanczos coefficients are stabilized but also when all the $z^{\lambda\mu}_{i,\mathbf{q}}$ coefficients reached their maxima and decay significantly.


\subsection{Sum rules}

In this section we present a validation of the sum rule given by Eq.~\eqref{eq:sum-rule-reciprocal_space}. Here we present the results not only for the CrI$_3$ monolayer, but also for bulk Fe and Ni, two prototypical metallic ferromagnets whose magnon dispersions have already been studied with the present algorithm and shown elsewhere~\cite{Gorni:2018}.

\begin{figure}[h!]
    \centering
    \includegraphics[width=0.95\textwidth,trim  = 0 20 0 20, clip]{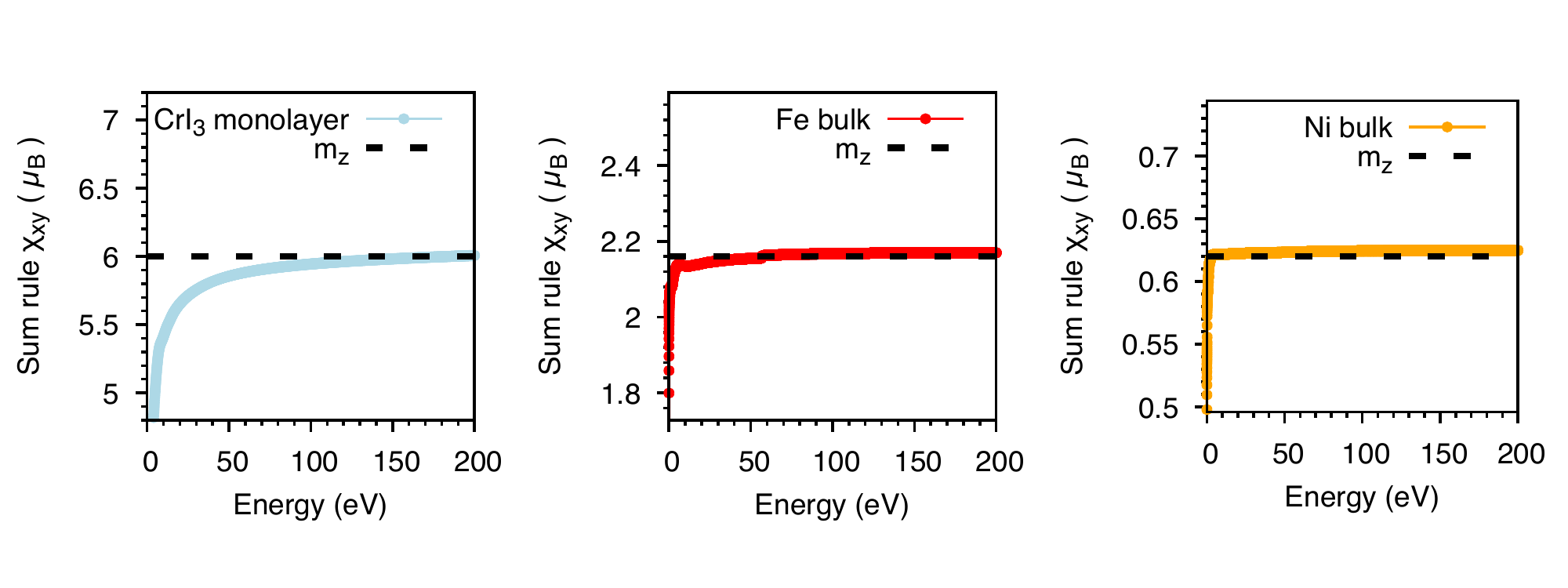}
    \caption{Sum rule check using Eq.~\eqref{eq:sum-rule-reciprocal_space} for the CrI$_3$ monolayer, bulk Fe, and bulk Ni. The transferred momentum is $\mathbf{q} = \mathrm{M}$ for CrI$_3$, $|\mathbf{q}| = 0.1 \, (2\pi /a)$ along the $\Gamma$-N direction for Fe~\cite{Gorni:2018}, and $|\mathbf{q}| = 0.1 \, (2\pi /a)$ along the $\Gamma$-X direction for Ni~\cite{Gorni:2018}. The value of the integral in Eq.~\eqref{eq:sum-rule-reciprocal_space} is reported as a function of the upper integration bound. Lorentzian broadening of $\eta=1.0$~meV was used for CrI$_3$, and $\eta = 10$~meV was used for Fe and Ni. $m_z = \langle \hat{m}_{z}\rangle$ is the $z$ component of the ground-state magnetization density as obtained from DFT.}
    \label{fig:sumrule}
\end{figure}

In practice, the upper bound in the integral in Eq.~\eqref{eq:sum-rule-reciprocal_space} must be replaced with some finite value of the frequency. We found that this upper bound is very large (several hundreds of eV) compared to typical energies of magnons (from tens to hundreds of meV). This is so because the integration of the spin susceptibility matrix takes into account also the Stoner excitations that appear at larger energies than magnons. As can be seen in Fig.~\ref{fig:sumrule}, the value of the integral converges to the value of the $z$ component of the DFT ground-state magnetization density with the accuracy of $\sim 1\%$ for all three systems. We checked that the sum rule is satisfied independently of the specific $\mathbf{q}$ point, the choice of the Lorentzian broadening, and the number of Lanczos iterations as shown in previous works~\cite{Baroni:2012,Timrov:2015}. Finally, we note that in the case of bulk Fe and Ni the Hamiltonian commutes with $\hat{S}_z$ and the additional relation $\chi_{xy}(\mathbf{q},\mathbf{q};\omega) = -\chi_{yx}(\mathbf{q},\mathbf{q};\omega)$ holds~\cite{DelRe:2021,Giuliani:2005}, so only one of the two components is needed for checking the sum rule using Eq.~\eqref{eq:sum-rule-reciprocal_space}.


\subsection{Scaling}

Scaling of the \turboMagnon{} code is one of the crucial aspects that requires a separate discussion which we present in this section. Here we analyze benchmark tests that were performed for the CrI$_3$ monolayer on the ``Galileo100'' HPC cluster at CINECA~\cite{CINECAlink}. Each node is equipped with 2 CPUs of type ``Intel CascadeLake 8260'', each CPU having 24 cores (2.4 GHz, 384 GB RAM); each node executes 24 MPI ranks, each one with 2 OpenMP threads. 

For the scaling tests we used a Monkhorst-Pack~\cite{Monkhorst:1976} $8\times 8\times 1$ $\mathbf{k}$ points mesh (we checked that the Monkhorst-Pack and $\Gamma$-centered $\mathbf{k}$ points meshes give the same converged spectra for the CrI$_3$ monolayer). The LL approach currently does not use symmetries, and hence in the ground-state DFT calculation we end up with the full grid of 64 points that contain $\mathbf{k}$ and $-\mathbf{k}$ (no inversion symmetry). In the linear-response calculation, we need to add to each of these points $\mathbf{q}$ and $-\mathbf{q}$, so that the final list of points is: $\mathbf{k}$, $-\mathbf{k}$, $\mathbf{k+q}$, $\mathbf{k-q}$, $\mathbf{-k+q}$, and $\mathbf{-k-q}$. Therefore, the total number of points that have to be considered when using the $8\times 8\times 1$ Monkhorst-Pack mesh is 192. It is important to note that e.g. for the hexagonal BZ (which is the case here) the $N \times N \times 1$ Monkhorst-Pack (i.e. shifted) $\mathbf{k}$ points meshes with $N$ being odd or for $N \times N \times 1$ $\Gamma$-centered (i.e. unshifted) $\mathbf{k}$ points meshes with $N$ being even some points fall on the edge of the BZ (for the $\Gamma$-centered $\mathbf{k}$ points meshes, the $\Gamma$ point always falls on the BZ border irrespective of $N$). These ``special $\mathbf{k}$ points'' are assigned a weight of $1/2$ and the $-\mathbf{k}$ point is generated, so the total number of points is larger than $3N^2$. 

\begin{figure}[t]
    \centering
    \includegraphics[width=\textwidth]{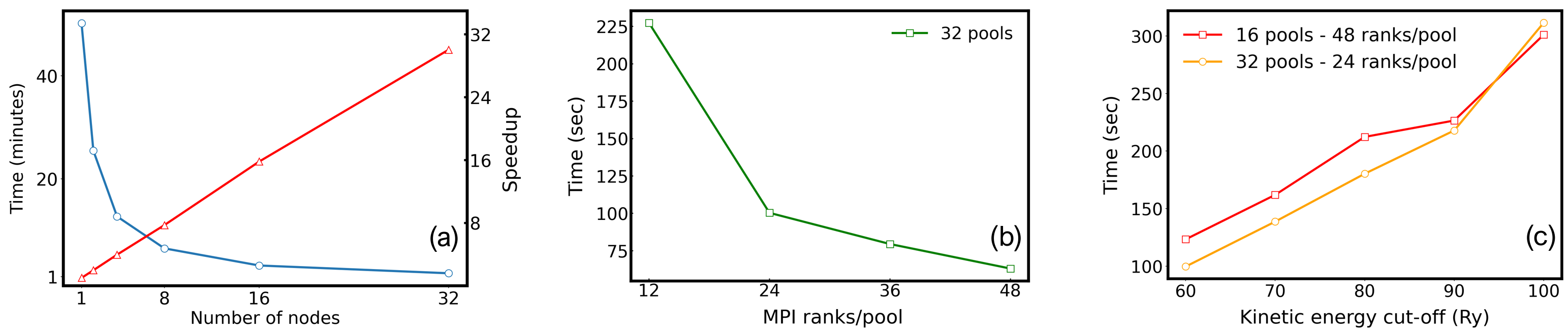}
    \caption{Scaling tests performed using the \turboMagnon{} code for the CrI$_3$ monolayer using a Monkhorst-Pack $8\times 8\times 1$ $\mathbf{k}$ points mesh, including SOC. (a)~Time per 100 Lanczos iterations (empty circles connected with blue lines, left axis) and the speedup (empty triangles connected with red lines, right axis) as a function of the number of compute nodes. The number of $\mathbf{k}$ points pools is equal to the number of compute nodes (i.e. the number of MPI ranks/pool is 24 for all data points). The kinetic-energy cutoff is 60~Ry for all data points. (b)~Time per 100 Lanczos iterations as a function of the number of MPI ranks/pool. The number of $\mathbf{k}$ points pools is equal to 32 and the kinetic-energy cutoff is 60~Ry for all data points. (c)~Time per 100 Lanczos iterations as a function of the kinetic-energy cutoff for wavefunctions using 32 compute nodes with two computational setups: 16 (empty squares connected with red lines) and 32 (empty circles connected with yellow lines) $\mathbf{k}$ points pools. The kinetic-energy cutoff for the charge and magnetization density and potentials is 4 times larger than the corresponding cutoffs for the wavefunctions since we use NC PPs.} 
    \label{fig:scaling}
\end{figure}

The overall scaling of the \turboMagnon{} code is obtained combining the plane-wave and $\mathbf{k}$ points parallelization levels. The TDDFpT calculations scale linearly with respect to the number of $\mathbf{k}$ and $\mathbf{q}$ points. We recall that all $\mathbf{q}$ points are independent and hence the TDDFpT calculations for different $\mathbf{q}$ points can be run independently and in parallel without communicating with each other. It is instructive to discuss the scaling of the \turboMagnon{} code using different number of compute nodes, number of $\mathbf{k}$ points pools, and number of MPI ranks per pool. These results are shown in Figs.~\ref{fig:scaling}~(a) and (b) that report time needed to perform 100 Lanczos iterations. In general, for the highest efficiency of computations the number of $\mathbf{k}$ points pools should be chosen such that the ratio between the number of pools and the number of nodes is an integer number. If this criteria is not satisfied and some pools are split between different nodes, the calculations are slower due to the loss of time for inter-node communications. For this reason, in Fig.~\ref{fig:scaling}~(a) the number of $\mathbf{k}$ points pools is chosen to be equal to the number of compute nodes (but of course the number of pools can be chosen to be larger than the number of nodes), and therefore the number of MPI ranks/pool is 24 for all data points. Moving forward, in this specific example we have 192 points, and these should be divided into pools in such a way that 6 ``sister points'' (i.e. $\mathbf{k}$, $-\mathbf{k}$, $\mathbf{k+q}$, $\mathbf{k-q}$, $\mathbf{-k+q}$, and $\mathbf{-k-q}$) reside in the same pool. For this reason, in Fig.~\ref{fig:scaling}~(a) we choose the number of nodes (pools) such that the ratio between 192 and the number of nodes (pools) is 6 times an integer number. By following this logic, the maximum number of nodes (pools) possible is 32 ($192/32=6$, i.e. 6 points per pool). Hence, in Fig.~\ref{fig:scaling}~(a) we see that the \turboMagnon{} code scales (quasi-)linearly with the number of compute nodes provided the number of $\mathbf{k}$ points pools equals the number of compute nodes: the speedup is $\sim 30$ when using 32 nodes compared to the calculation using only 1 node. We note in passing that it is possible to use the number of nodes even larger than 32 (e.g. 48, 64, etc.) however in this case the number of MPI tasks/pool becomes larger than 24. This brings us to Fig.~\ref{fig:scaling}~(b) where we investigate the scaling of the \turboMagnon{} code as a function of the MPI ranks/pool. In order to exploit an increasingly large number of processors maintaining a high parallel efficiency it is necessary to tune the number of MPI ranks used in each pool. In order to investigate this point, we fixed the number of $\mathbf{k}$ points pools to 32 and changed the number of compute nodes (16, 32, 48, and 64). We can see in Fig.~\ref{fig:scaling}~(b) that the time needed to perform 100 Lanczos iterations decreases when we increase the number of MPI ranks/pool as expected, and that there is a change in the slope at 24 MPI ranks/pool. This change in the slope indicates that the parallel efficiency drops when the number of MPI ranks/pool is larger than 24. The trade-off between the total amount of computational time needed to solve the TDDFpT equations and the speedup has to be found for large-scale production calculations on HPCs - from Fig.~\ref{fig:scaling}~(a) and (b) we can conclude that this is achieved when using 32 nodes and 32 $\mathbf{k}$ points pools for the current example (i.e. 24 MPI ranks/pool).

In contrast, the scaling of the \turboMagnon{} code with respect to the number of atoms in the system is cubic. Finally, in Fig.~\ref{fig:scaling}~(c) we show the scaling with respect to the kinetic-energy cutoff for wavefunctions $E_\mathrm{cut}$ (the kinetic-energy cutoff for the charge and magnetization density and potentials is 4 times larger since we use NC PPs). For these tests we used 32 compute nodes and different number of pools, 16 and 32 (in order to see trends with respect to a different number of MPI ranks/pool). We see that in both cases when increasing $E_\mathrm{cut}$ from 60 to 100~Ry (i.e. by 40\%) the computational time is increased by a factor of $\sim 3$ (i.e. by 300\%). Interestingly, when using 32 $\mathbf{k}$ points pools we see a straight line between 60 and 90~Ry and then a change in the slope, while when using 16 $\mathbf{k}$ points pools the data is somewhat more noisy. Overall, we can conclude that the computational cost of \turboMagnon{} calculations increases rapidly when increasing $E_\mathrm{cut}$ and thus it is important to optimize the value of $E_\mathrm{cut}$ by checking the converge of the magnetic spectra.

To conclude, we use the scaling tests discussed above to highlight the computational cost for magnetic spectra calculations for the CrI$_3$ monolayer. Using the Monkhorst-Pack $8\times 8\times 1$ $\mathbf{k}$ points mesh and $E_\mathrm{cut}=60$~Ry we choose the optimal setup of 32 compute nodes and 32 $\mathbf{k}$ points pools (i.e. 24 MPI ranks/pool). With this setting, it takes 1 hour 40 minutes to perform 10000 Lanczos iterations (with SOC), which means approximately 53 node hours. This result does not depend on the value of the $\mathbf{q}$ point since we do not use symmetries; however, as we discussed above the number of Lanczos iterations needed to converge the spectra varies for different $\mathbf{q}$ points. Finally, without SOC the computational cost is slightly smaller but not substantially (since calculations were also performed in the noncollinear framework). 

\section{Conclusions}
\label{sec:Conclusions}

We have presented the \turboMagnon code as a component of the \QE\, distribution that implements a Liouville-Lanczos approach to time-dependent density-functional perturbation theory for the computation of magnetic spectra for any transferred momentum $\mathbf{q}$. The \turboMagnon code does not require the calculation of empty electronic states due to the use of standard techniques of the static density-functional perturbation theory~\cite{Baroni:2001}. The solution of the linear-response equations is done in the frequency domain using the non-Hermitian or pseudo-Hermitian Lanczos recursive algorithms that allow to avoid computationally expensive inversions of response matrices. The \turboMagnon code is implemented in the noncolinear spin-polarized framework including relativistic effects such as spin-orbit coupling. The effectiveness of the code is showcased on the example of the 2D ferromagnetic insulator CrI$_3$ and the fulfillment of linear-response sum rules is verified for monolayer CrI$_3$, bulk Fe and bulk Ni.

In the same spirit as the \QE\, project, \turboMagnon provides scientists worldwide with a well commented and open-source framework for implementing their ideas. It is in our best hope that \turboMagnon can benefit from the already well established users community of \QE\, for incorporating new ideas and keep growing in the future. The \turboMagnon code is hosted in a community accessible Git repository~\cite{QuantumESPRESSO:Gitlab} and hence, apart from the releases of \QE~\cite{QuantumESPRESSO:website}, researchers who are willing to test the latest experimental implementations are welcome to do so and to contribute with their feedback.

The \turboMagnon code can be extended so as to employ more advanced exchange-correlation functionals (e.g. to include the Hubbard $U$ correction or SCAN meta-GGA~\cite{Sun:2015}), to use ultrasoft and projector-augmented-wave pseudopotentials, use symmetry, to name a few. Moreover, one of our core goals is to generalize the \turboMagnon code to make it run on modern GPU architectures.

\section*{Acknowledgements}

We thank Andrea Dal Corso for fruitful discussions. This work was partially funded by the European Union through the \textsc{MaX} Centre of Excellence for Supercomputing applications (Project No. 824143), by the Italian MIUR/MUR through the PRIN 2017 \emph{FERMAT} grant and by the Swiss National Science Foundation (SNSF), through grant 200021-179138, and its National Centre of Competence in Research (NCCR) MARVEL. Computer time was provided by CINECA.

\appendix


\section{Sample input files}
\label{sec:Sample_input_files}

\vskip 0.5 cm

\noindent
{\bf Input example 1:} Input sample for \texttt{pw.x}
\begin{verbatim}
&control
    calculation = 'scf'
    restart_mode = 'from_scratch'
    prefix = 'CrI3'
    pseudo_dir = './pseudo'
    outdir = './tmp'
 /
 &system
    ibrav = 4
    celldm(1) = 12.98
    celldm(3) =  2.88
    nat = 8
    ntyp = 2
    ecutwfc = 60.0
    occupations = 'smearing'
    smearing = 'gauss'
    degauss = 0.01
    starting_magnetization(1) = 0.5
    lspinorb = .true.
    noncolin = .true.
    nosym = .true.
    noinv = .true.
 /
 &electrons
    diagonalization = 'david'
    mixing_mode   = 'plain'
    mixing_beta = 0.3
    conv_thr =  1.d-13
 /
ATOMIC_SPECIES
Cr     51.996    Cr.rel-pz-n-nc.UPF
I     126.9045   I.rel-pz-n-nc.UPF
ATOMIC_POSITIONS {crystal}
Cr  0.6666667  0.3333333  0.6667240
Cr  0.0000000  0.0000000  0.6666093
I   0.9788986  0.6665119  0.7464249
I   0.6876134  0.0211013  0.7464249
I   0.3334880  0.3123867  0.7464249
I   0.6877680  0.6668213  0.5869084
I   0.9790533  0.3122319  0.5869084
I   0.3331787  0.0209467  0.5869084
K_POINTS {automatic}
8 8 1 0 0 0
\end{verbatim}

\vskip 0.5 cm

\noindent
{\bf Input example 2:} Input sample for \texttt{turbo\_magnon.x}
\begin{verbatim}
&lr_input
    prefix = 'CrI3'
    outdir = './tmp'
    restart_step = 200
    restart = .false.
/
&lr_control
    itermax = 15000
    pseudo_hermitian = .true.
    q1 = 0.000
    q2 = 0.577
    q3 = 0.000
    ipol = 4
/
\end{verbatim}

\vskip 0.5 cm

\noindent
{\bf Input example 3:} Input sample for \texttt{turbo\_spectrum.x}
\begin{verbatim}
&lr_input
    prefix = 'CrI3'
    outdir = './tmp'
    magnons = .true.
    itermax0 = 15000
    itermax = 15000
    extrapolation = 'no'
    ipol = 4
    units = 3
    epsil = 1.0
    start = 0.0
    end = 100.0
    increment = 0.1
/\end{verbatim}

\section{Input Variables}
\label{sec:Input_variables}


\begin{table}[H]
 \begin{tabular}{|>{\centering}m{0.6cm}|c}
      \cline{1-1} 
      Card  & 
      \begin{tabular}{|>{\centering}m{3.45cm}|>{\centering}m{1.5cm}|>{\centering}p{9.0cm}|} \hline 
      Variable name  & Default Value  & Description \tabularnewline
      \end{tabular} \tabularnewline
      \cline{1-1} 
\begin{sideways}
   \textbf{lr\_input} 
\end{sideways}
      & \begin{tabular}{|>{\raggedright}m{3.45cm}|> {\centering}m{1.5cm}|m{9.0cm}|} \hline 
        \texttt{prefix} & '\texttt{pwscf}' & {\footnotesize 
          The files generated by the ground state \texttt{pw.x} run should have this same prefix.}\tabularnewline 
        \texttt{outdir} & './' & {\footnotesize
          Working directory. On start, it should contain the files generated by a ground state \texttt{pw.x} run.}\tabularnewline 
        \texttt{restart} & \emph{.false.} & {\footnotesize 
          When set to \emph{.true.}, \texttt{turbo\_magnon.x} will attempt to restart from a previous interrupted calculation (see \texttt{restart\_step} variable).} \tabularnewline
        \texttt{restart\_step} & \texttt{itermax} & {\footnotesize 
          The code writes restart files every \texttt{restart\_step} iterations. Restart files are automatically written at the end of
          \texttt{itermax} Lanczos steps.} \tabularnewline 
        \texttt{lr\_verbosity} & 1 & {\footnotesize Verbosity level: the larger the value the more data is printed in the output file.} \tabularnewline 
      \end{tabular} \tabularnewline
      \cline{1-1} 
\begin{sideways}
  \textbf{lr\_control}  
\end{sideways}
      & \begin{tabular}{|>{\raggedright}m{3.45cm}|>{\centering}m{1.5cm}|m{9.0cm}|} \hline 
        \texttt{itermax} & 500 & {\footnotesize Number of Lanczos iterations to be performed.} \tabularnewline 
        \texttt{q1, q2, q3}    & 1, 1, 1 & {\footnotesize Cartesian components of the transferred momentum $\mathbf{q}$ in units of $2 \pi/a$ (where $a$ is the lattice parameter of the unit cell).} \tabularnewline 
        \texttt{pseudo\_hermitian} & \emph{.true.} & \footnotesize{If \emph{.true.} then the pseudo-Hermitian Lanczos algorithm is used, if \emph{.false.} then the non-Hermitian Lanczos biorthogonalization algorithm is used (which is two times slower). 
          } \tabularnewline 
        \texttt{ipol} & 1 & \footnotesize{Polarization direction of the magnetic field (\texttt{ipol=1} for $x$ direction, \texttt{ipol=2} for $y$ direction,  \texttt{ipol=3} for $z$ direction). \texttt{ipol} defines the column of the susceptibility tensor $\chi_{\lambda\mu}(\mathbf{q},\mathbf{q};\omega)$ to be computed, which corresponds to one Lanczos chain. \texttt{ipol=4} computes the full susceptibility tensor (3 Lanczos chains).} \tabularnewline  
         \hline
      \end{tabular} \tabularnewline
      \cline{1-1}
 \end{tabular} 
 \caption{Input variables for \texttt{turbo\_magnon.x} }
 \label{tab:Table_input_turbo_magnon.x}
\end{table}


\begin{table}[H]
 \begin{tabular}{|>{\centering}m{0.6cm}|c}
      \cline{1-1}
      Card  &
      \begin{tabular}{|>{\centering}m{3.45cm}|>{\centering}m{1.5cm}|>{\centering}p{9.0cm}|} \hline
      Variable name  & Default Value  & Description \tabularnewline
      \end{tabular} \tabularnewline
      \cline{1-1}
\begin{sideways}
   \textbf{lr\_input}
\end{sideways}
      & \begin{tabular}{|>{\raggedright}m{3.45cm}|> {\centering}m{1.5cm}|m{9.0cm}|} \hline
        \texttt{prefix} & '\texttt{pwscf}' & {\footnotesize
          Prefix of the files generated by the previous \texttt{turbo\_magnon.x} run.}\tabularnewline
        \texttt{outdir} & './' & {\footnotesize
          The directory where the output files produced by the previous \texttt{turbo\_magnon.x} run are stored.}\tabularnewline
        \texttt{magnons} & \emph{.false.} & {\footnotesize
          Must be set to \emph{.true.} for the magnetic spectrum calculation.}\tabularnewline
        \texttt{itermax0} & 1000 & {\footnotesize
          Number of Lanczos coefficients to be read from the file.}\tabularnewline
        \texttt{itermax} & 1000 & {\footnotesize
          The total number of Lanczos coefficients that will be considered in the calculation of the spin susceptibility matrix. If \texttt{itermax} $>$ \texttt{itermax0}, the Lanczos coefficients in between 
          \texttt{itermax0}+1 and \texttt{itermax} will be extrapolated.}\tabularnewline
        \texttt{extrapolation} & '\texttt{no}' & {\footnotesize
          Sets the extrapolation scheme for Lanczos coefficients. '\texttt{osc}' = bi-constant extrapolation; 
          '\texttt{constant}' = constant extrapolation; '\texttt{no}' = no extrapolation.}\tabularnewline
        \texttt{ipol} & 1 & {\footnotesize
          Same meaning of the \texttt{ipol} variable as in the \texttt{turbo\_magnon.x} input.
          }\tabularnewline
        \texttt{units} & 0 & {\footnotesize
          Units for \texttt{epsil}, \texttt{start}, \texttt{end}, and \texttt{increment}. 0 = Ry, 1 = eV, 2 = nm, 3 = meV. Only \texttt{units}=3 is allowed for \texttt{magnons}=\emph{.true.}}\tabularnewline  
          \texttt{epsil} & 0.02 & {\footnotesize
          The Lorentzian broadening parameter $\eta$ (in \texttt{units}).}\tabularnewline
        \texttt{start} & 0.0 & {\footnotesize
          The susceptibility is computed starting from this value of $\omega$ (in \texttt{units}).
          }\tabularnewline
        \texttt{end} & 2.5 & {\footnotesize
          The susceptibility is computed up to this value of $\omega$ (in \texttt{units}).
          }\tabularnewline
        \texttt{increment} & 0.001 & {\footnotesize
          Incremental step $\Delta \omega$ used to define the mesh between \texttt{start} and \texttt{end} (in \texttt{units}).}\tabularnewline
        \texttt{verbosity} & 0 & {\footnotesize 
          Verbosity level: the larger the value the more data is printed in the output file.} \tabularnewline \hline
      \end{tabular} \tabularnewline
      \cline{1-1}
 \end{tabular}
 \caption{Input variables for \texttt{turbo\_spectrum.x} }
 \label{tab:Table_input_turbo_spectrum.x}
\end{table}


\begin{thebibliography}{100}
\expandafter\ifx\csname url\endcsname\relax
  \def\url#1{\texttt{#1}}\fi
\expandafter\ifx\csname urlprefix\endcsname\relax\def\urlprefix{URL }\fi
\expandafter\ifx\csname href\endcsname\relax
  \def\href#1#2{#2} \def\path#1{#1}\fi

\bibitem{Mook:1973}
H.~Mook, R.~Nicklow, Phys. Rev. B 7 (1973) 336.

\bibitem{Qin:2015}
H.~Qin, K.~Zakeri, A.~Ernst, L.~Sandratskii, P.~Buczek, A.~Marmodoro, T.-H.
  Chuang, Y.~Zhang, J.~Kirschner, Long-living terahertz magnons in ultrathin
  metallic ferromagnets, Nat. Commun. 6 (2015) 6126.

\bibitem{Hirjibehedin:2006}
C.~Hirjibehedin, J.~Lutz, A.~Heinrich, Science 312 (2006) 1021.

\bibitem{Chaix:2018}
L.~Chaix, E.~W. Huang, S.~Gerber, X.~Lu, C.~Jia, Y.~Huang, D.~E. McNally,
  Y.~Wang, F.~H. Vernay, A.~Keren, M.~Shi, B.~Moritz, Z.-X. Shen, T.~Schmitt,
  T.~P. Devereaux, W.-S. Lee, {Resonant inelastic x-ray scattering studies of
  magnons and bimagnons in the lightly doped cuprate
  ${\mathrm{La}}_{2\ensuremath{-}x}{\mathrm{Sr}}_{x}{\mathrm{CuO}}_{4}$}, Phys.
  Rev. B 97 (2018) 155144.

\bibitem{Brookes:2020}
N.~Brookes, D.~Betto, K.~Cao, Y.~Lu, K.~Kummer, F.~Giustino, Spin waves in
  metallic iron and nickel measured by soft x-ray resonant inelastic
  scattering, Physical Review B 102~(6) (2020) 064412.

\bibitem{Lebert:2020}
B.~W. Lebert, S.~Kim, V.~Bisogni, I.~Jarrige, A.~M. Barbour, Y.-J. Kim,
  Resonant inelastic x-ray scattering study of-rucl3: a progress report,
  Journal of Physics: Condensed Matter 32~(14) (2020) 144001.

\bibitem{Pellicciari:2021}
J.~Pelliciari, S.~Karakuzu, Q.~Song, R.~Arpaia, A.~Nag, M.~Rossi, J.~Li, T.~Yu,
  X.~Chen, R.~Peng, et~al., Evolution of spin excitations from bulk to
  monolayer fese, Nature communications 12~(1) (2021) 1--8.

\bibitem{Pelliciari:2021aa}
J.~Pelliciari, S.~Lee, K.~Gilmore, J.~Li, Y.~Gu, A.~Barbour, I.~Jarrige, C.~H.
  Ahn, F.~J. Walker, V.~Bisogni, Tuning spin excitations in magnetic films by
  confinement, Nature Materials 20~(2) (2021) 188--193.

\bibitem{Costa:2010}
A.~Costa, R.~Muniz, S.~Lounis, A.~Klautau, D.~Mills, Spin-orbit coupling and
  spin waves in ultrathin ferromagnets: The spin-wave rashba effect, Phys. Rev.
  B 82 (2010) 014428.

\bibitem{Bergman:2010}
A.~Bergman, A.~Taroni, L.~Bergqvist, J.~Hellsvik, B.~Hj\"orvarsson,
  O.~Eriksson, Magnon softening in a ferromagnetic monolayer: A
  first-principles spin dynamics study, Phys. Rev. B 81 (2010) 144416.

\bibitem{Zakeri:2012}
K.~Zakeri, Y.~Zhang, T.-H. Chuang, J.~Kirschner, Magnon lifetimes on the
  fe(110) surface: The role of spin-orbit coupling, Phys. Rev. Lett. 108 (2012)
  197205.

\bibitem{Zakeri:2017}
K.~Zakeri, Probing of the interfacial heisenberg and dzyaloshinskii–moriya
  exchange interaction by magnon spectroscopy, J. Phys.: Condens. Matter 29
  (2017) 013001.

\bibitem{Savrasov:1998}
S.~Savrasov, Linear response calculations of spin fluctuations, Phys. Rev.
  Lett. 81 (1998) 2570.

\bibitem{Lounis:2011}
S.~Lounis, A.~Costa, R.~Muniz, D.~Mills, Phys. Rev. B 83 (2011) 035109.

\bibitem{Buczek:2011b}
P.~Buczek, A.~Ernst, L.~Sandratskii, Different dimensionality trends in the
  landau damping of magnons in iron, cobalt, and nickel: Time-dependent density
  functional study, Phys. Rev. B 84 (2011) 174418.

\bibitem{Rousseau:2012}
B.~Rousseau, A.~Eiguren, A.~Bergara, Efficient computation of magnon
  dispersions within time-dependent density functional theory using maximally
  localized wannier functions, Phys. Rev. B 85 (2012) 054305.

\bibitem{dosSantosDias:2015}
M.~{dos}~{Santos}~{D}ias, B.~Schweflinghaus, S.~Bl\"ugel, S.~Lounis,
  Relativistic dynamical spin excitations of magnetic adatoms, Phys. Rev. B 91
  (2015) 075405.

\bibitem{Wysocki:2017}
A.~Wysocki, V.~Valmispild, A.~Kutepov, S.~Sharma, J.~Dewhurst, E.~Gross,
  A.~Lichtenstein, V.~Antropov, Spin-density fluctuations and the
  fluctuation-dissipation theorem in 3d ferromagnetic metals, Phys. Rev. B 96
  (2017) 184418.

\bibitem{Cao:2018}
K.~Cao, H.~Lambert, P.~Radaelli, F.~Giustino, Ab initio calculation of spin
  fluctuation spectra using time-dependent density functional perturbation
  theory, plane waves, and pseudopotentials, Phys. Rev. B 97 (2018) 024420.

\bibitem{TancogneDejean:2020}
N.~Tancogne-Dejean, F.~Eich, A.~Rubio, {Time-Dependent Magnons from First
  Principles}, J. Chem. Theory Comput. 16 (2020) 1007.

\bibitem{Skovhus:2021}
T.~Skovhus, T.~Olsen, {Dynamic transverse magnetic susceptibility in the
  projector augmented-wave method: Application to Fe, Ni, and Co}, Phys. Rev. B
  103 (2021) 245110.

\bibitem{Aryasetiawan:1999}
F.~Aryasetiawan, K.~Karlsson, Phys. Rev. B 60 (1999) 7419.

\bibitem{Karlsson:2000}
K.~Karlsson, F.~Aryasetiawan, Phys. Rev. B 62 (2000) 3006.

\bibitem{Kotani:2008}
T.~Kotani, M.~van {S}chilfgaarde, J. Phys.: Condens. Matter 20 (2008) 295214.

\bibitem{Sasioglu:2010}
\c{S}a\c{s}io\u{g}lu, A.~Schindlmayr, C.~Friedrich, F.~Freimuth, S.Bl\"ugel,
  Phys. Rev. B 81 (2010) 054434.

\bibitem{Muller:2016}
M.~M\"uller, C.~Friedrich, S.~Bl\"ugel, Phys. Rev. B 94 (2016) 064433.

\bibitem{Runge:1984}
E.~Runge, E.~Gross, Phys. Rev. Lett. 52 (1984) 997.

\bibitem{Marques:2012}
M.~A.~L. Marques, N.~T. Maitra, F.~M.~S. Nogueira, E.~K.~U. Gross, A.~Rubio
  (Eds.), {F}undamentals of {T}ime-{D}ependent {D}ensity {F}unctional {T}heory,
  Vol. 837, {L}ecture {N}otes in {Physics}, {S}pringer-{V}erlag, Berlin
  Heidelberg, 2012.

\bibitem{Baroni:2012}
S.~Baroni, R.~Gebauer, The Liouville-Lanczos {A}pproach to {T}ime-{D}ependent
  {D}ensity-{F}unctional ({P}erturbation) {Theory}, Ref.~\cite{Marques:2012},
  chapter 19, p.~375-390.

\bibitem{Rocca:2008}
D.~Rocca, R.~Gebauer, Y.~Saad, S.~Baroni, J. Chem. Phys. 128 (2008) 154105.

\bibitem{Timrov:2013}
I.~Timrov, N.~Vast, R.~Gebauer, S.~Baroni, Phys. Rev. B 88 (2013) 064301, {\it
  ibid.} {\bf 91}, 139901 (2015).

\bibitem{Baroni:1987}
S.~Baroni, P.~Giannozzi, A.~Testa, {G}reen's-function approach to linear
  response in solids, Phys. Rev. Lett. 58 (1987) 1861.

\bibitem{Baroni:2001}
S.~Baroni, S.~de~Gironcoli, A.~D. Corso, P.~Giannozzi, Phonons and related
  crystal properties from density-functional perturbation theory, Rev. Mod.
  Phys. 73~(2) (2001) 515.

\bibitem{Malcioglu:2011}
O.~Malcioi\u{g}lu, R.~Gebauer, D.~Rocca, S.~Baroni, Comput. Phys. Commun. 182
  (2011) 1744.

\bibitem{Ge:2014}
X.~Ge, S.~J. Binnie, D.~Rocca, R.~Gebauer, S.~Baroni, turbo{TDDFT} 2.0 --
  {H}ybrid functionals and new algorithms within time-dependent
  density-functional perturbation theory, Comput. Phys. Commun. 185 (2014)
  2080.

\bibitem{Timrov:2015}
I.~Timrov, N.~Vast, R.~Gebauer, S.~Baroni, Comput. Phys. Commun. 196 (2015)
  460.

\bibitem{Timrov:2017}
I.~Timrov, M.~Markov, T.~Gorni, M.~Raynaud, O.~Motornyi, R.~Gebauer, S.~Baroni,
  N.~Vast, Phys. Rev. B 95 (2017) 094301.

\bibitem{Motornyi:2020}
O.~Motornyi, N.~Vast, I.~Timrov, O.~Baseggio, S.~Baroni, A.~{Dal Corso}, Phys.
  Rev. B 102 (2020) 035156.

\bibitem{GPL}
The GNU General Public License: \url{http://www.gnu.org/licenses/gpl.html}.

\bibitem{Giannozzi:2009}
P.~Giannozzi, S.~Baroni, N.~Bonini, M.~Calandra, R.~Car, C.~Cavazzoni,
  D.~Ceresoli, G.~Chiarotti, M.~Cococcioni, I.~Dabo, A.~Dal~Corso,
  S.~De~Gironcoli, S.~Fabris, G.~Fratesi, R.~Gebauer, U.~Gerstmann,
  C.~Gougoussis, A.~Kokalj, M.~Lazzeri, L.~Martin-Samos, N.~Marzari, F.~Mauri,
  R.~Mazzarello, S.~Paolini, A.~Pasquarello, L.~Paulatto, C.~Sbraccia,
  S.~Scandolo, G.~Sclauzero, A.~Seitsonen, A.~Smogunov, P.~Umari,
  R.~Wentzcovitch, {Q}uantum {ESPRESSO}: {A} modular and open-source software
  project for quantum simulations of materials, J. Phys.: Condens. Matter 21
  (2009) 395502.

\bibitem{Giannozzi:2017}
P.~Giannozzi, O.~Andreussi, T.~Brumme, O.~Bunau, M.~Buongiorno~Nardelli,
  M.~Calandra, R.~Car, C.~Cavazzoni, D.~Ceresoli, M.~Cococcioni, N.~Colonna,
  I.~Carnimeo, A.~Dal~Corso, S.~de~Gironcoli, P.~Delugas, R.~A.
  Di{S}tasio~{J}r., A.~Ferretti, A.~Floris, G.~Fratesi, G.~Fugallo, R.~Gebauer,
  U.~Gerstmann, F.~Giustino, T.~Gorni, J.~Jia, M.~Kawamura, H.-Y. Ko,
  A.~Kokalj, E.~K\"{u}\c{c}\"{u}kbenli, M.~Lazzeri, M.~Marsili, N.~Marzari,
  F.~Mauri, N.~L. Nguyen, H.-V. Nguyen, A.~Otero-de-la {R}osa, L.~Paulatto,
  S.~Ponc\'e, D.~Rocca, R.~Sabatini, B.~Santra, M.~Schlipf, A.~Seitsonen,
  A.~Smogunov, I.~Timrov, T.~Thonhauser, P.~Umari, N.~Vast, S.~Baroni,
  {A}dvanced capabilities for materials modelling with {Q}uantum {ESPRESSO}, J.
  Phys.: Condens. Matter 29 (2017) 465901.

\bibitem{Giannozzi:2020}
P.~Giannozzi, O.~Baseggio, P.~Bonf\`a, D.~Brunato, R.~Car, I.~Carnimeo,
  C.~Cavazzoni, S.~de~Gironcoli, P.~Delugas, F.~Ferrari~Ruffino, A.~Ferretti,
  N.~Marzari, I.~Timrov, A.~Urru, S.~Baroni, {Quantum ESPRESSO toward the
  exascale}, J. Chem. Phys. 152 (2020) 154105.

\bibitem{Halpern:1939}
O.~Halpern, M.~Johnson, {On the Magnetic Scattering of Neutrons}, Phys. Rev. 55
  (1939) 898.

\bibitem{Blume:1963}
M.~Blume, {Polarization Effects in the Magnetic Elastic Scattering of Slow
  Neutrons}, Phys. Rev. 130 (1963) 1670.

\bibitem{Jones:1985}
W.~Jones, N.~H. March, Theoretical solid state physics, Volume 1, Courier
  Corporation, 1985.

\bibitem{Gorni:2018}
T.~Gorni, I.~Timrov, S.~Baroni, Spin dynamics from time-dependent density
  functional perturbation theory, Eur. Phys. J. B 91 (2018) 249.

\bibitem{Gorni:2016}
T.~Gorni, {S}pin-fluctuation spectra in magnetic systems: a novel approach
  based on tddft, Ph.D. thesis, Scuola Internazionale Superiore di Studi
  Avanzati (SISSA), Trieste, Italy (2016,
  \url{http://hdl.handle.net/20.500.11767/43342}).

\bibitem{Notations}
We use a hat ``$\hat{\phantom{H}}$'' on top of letters to indicate operators
  (e.g. $\hat{V}$), while these same operators in the coordinate representation
  are written without the hat and with the explicit dependence on the position
  vector $\mathbf{r}$ [e.g. $V(\mathbf{r})$]. Moreover, we use a tilde
  ``$\tilde{\phantom{V}}$'' to indicate a Fourier transform of various
  quantities from the time domain [e.g. $V(t)$] to the frequency domain [e.g.
  $\tilde{V}(\omega)$]. A combination of these two notations is often used in
  this work.

\bibitem{Pot_notation}
With the upper case letter we denote a $2 \times 2$ matrix potential, while
  with the lower case letter we denote scalar potentials.

\bibitem{Kleinman:1980}
L.~Kleinman, Phys. Rev. B 21 (1980) 2630.

\bibitem{Bachelet:1982a}
G.~Bachelet, Schl\"uter, Phys. Rev. B 25 (1982) 2103.

\bibitem{Bachelet:1982b}
G.~Bachelet, D.~Hamann, Schl\"uter, Phys. Rev. B 26 (1982) 4199.

\bibitem{Hemstreet:1993}
L.~Hemstreet, C.~Fong, J.~Nelson, Phys. Rev. B 47 (1993) 4238.

\bibitem{Ceresoli:2010}
D.~Ceresoli, U.~Gerstmann, A.~Seitsonen, F.~Mauri, Phys. Rev. B 81 (2010)
  060409(R).

\bibitem{Ashcroft:1976}
N.~Ashcroft, N.~Mermin, Solid State Physics, Saunders College Publishing,
  Philadelphia, 1976.

\bibitem{DalCorso:2010}
A.~{Dal~Corso}, Phys. Rev. B 82 (2010) 075116.

\bibitem{Note:notation_Vxc}
{W}e note that in the second term of Eq.~\eqref{eq:v_XC_q} we symbolically mean
  a scalar product, while in the second term of Eq.~\eqref{eq:b_XC_q} we
  symbolically mean a matrix-vector multiplication.

\bibitem{Note:density_property}
{T}hese relations are a consequence of the fact that charge- and
  magnetization-density responses are real functions in space and time.

\bibitem{Singh:2019}
N.~Singh, P.~Elliott, T.~Nautiyal, J.~K. Dewhurst, S.~Sharma, Phys. Rev. B 99
  (2019) 035151.

\bibitem{Lehtola:2022}
S.~Lehtola, M.~Marques, {Many recent density functionals are numerically
  unstable}, arXiv:2206.14062 (2022).

\bibitem{Sun:2015}
J.~Sun, A.~Ruzsinszky, J.~Perdew, Phys. Rev. Lett. 115 (2015) 036402.

\bibitem{Ekholm:2018}
M.~Ekholm, D.~Gambino, H.~J\"onsson, F.~Tasn\'adi, B.~Alling, I.~Abrikosov,
  {Assessing the SCAN functional for itinerant electron ferromagnets}, Phys.
  Rev. B 98 (2018) 094413.

\bibitem{Tran:2020}
F.~Tran, G.~Baudesson, J.~Carrete, G.~Madsen, P.~Blaha, K.~Schwarz, D.~Singh,
  {Shortcomings of meta-GGA functionals when describing magnetism}, Phys. Rev.
  B 102 (2020) 024407.

\bibitem{Skovhus:2022}
T.~Skovhus, T.~Olsen, H.~Ronnow, arXiv:2110.07282 (2022).

\bibitem{Skovhus:2022b}
T.~Skovhus, T.~Olsen, arXiv:2203.04796 (2022).

\bibitem{Gokhale:1992}
M.~Gokhale, A.~Ormeci, D.~Mills, Phys. Rev. B 46 (1992) 8978.

\bibitem{Saad:2003}
Y.~Saad, Iterative Methods for Sparse Linear Systems, 2nd Edition, SIAM,
  Philadelphia, 2003.

\bibitem{Gruning:2011}
M.~Gr\"uning, A.~Marini, X.~Gonze, Comput. Math. Sci. 50 (2011) 2148.

\bibitem{Mostafazadeh:2002}
A.~Mostafazadeh, {Pseudo-Hermiticity versus PT symmetry: The necessary
  condition for the reality of the spectrum of a non-Hermitian Hamiltonian}, J.
  Math. Phys. 43 (2002) 205.

\bibitem{GNUautoconf}
{GNU Autoconf: \url{https://www.gnu.org/software/autoconf}}.

\bibitem{cmake}
CMake is an open-source, cross-platform family of tools designed to build, test
  and package software: \url{https://cmake.org/}.

\bibitem{MPI:1994}
Message passing interface forum, Int. J. Supercomput. Appl. 8 (1994) 159.

\bibitem{THEOSlib}
\url{http://theossrv1.epfl.ch/Main/Pseudopotentials}.

\bibitem{Soriano:2020}
D.~Soriano, M.~I. Katsnelson, J.~Fern{\'a}ndez-Rossier, Magnetic
  two-dimensional chromium trihalides: A theoretical perspective, Nano Letters
  20~(9) (2020) 6225--6234.

\bibitem{MaterialsCloudArchive2022}
T.~Gorni, O.~Baseggio, P.~Delugas, S.~Baroni, I.~Timrov,
  \href{https://archive.materialscloud.org/record/2022.89}{{\texttt{turboMagnon}
  -- A code for the simulation of spin-wave spectra using the Liouville-Lanczos
  approach to time-dependent density-functional perturbation theory}},
  Materials Cloud Archive \textbf{2022.89} (2022),
  doi:~10.24435/materialscloud:6j-kd.
\newblock \href {https://doi.org/10.24435/materialscloud:6j-kd}
  {\path{doi:10.24435/materialscloud:6j-kd}}.
\newline\urlprefix\url{https://archive.materialscloud.org/record/2022.89}

\bibitem{Goldstone:1962}
J.~Goldstone, A.~Salam, S.~Weinberg, Broken symmetries, Phys. Rev. 127 (1962)
  965--970.

\bibitem{Watanabe:2012}
H.~Watanabe, H.~Murayama, Unified description of nambu-goldstone bosons without
  lorentz invariance, Phys. Rev. Lett. 108 (2012) 251602.

\bibitem{Delugas:2021}
P.~Delugas, O.~Baseggio, I.~Timrov, S.~Baroni, T.~Gorni, {Magnon-phonon
  interactions open a gap at the Dirac point in the spin-wave spectra of
  CrI$_3$ 2D magnets}, submitted (arXiv:2203.01120) (2021).

\bibitem{Chen:2018}
L.~Chen, J.-H. Chung, B.~Gao, T.~Chen, M.~B. Stone, A.~I. Kolesnikov, Q.~Huang,
  P.~Dai, {Topological spin excitations in honeycomb ferromagnet CrI$_3$},
  Phys. Rev. X 8~(4) (2018) 041028.

\bibitem{Chen:2020}
L.~Chen, J.-H. Chung, T.~Chen, C.~Duan, A.~Schneidewind, I.~Radelytskyi, D.~J.
  Voneshen, R.~A. Ewings, M.~B. Stone, A.~I. Kolesnikov, B.~Winn, S.~Chi, R.~A.
  Mole, D.~H. Yu, B.~Gao, P.~Dai, {Magnetic anisotropy in ferromagnetic
  ${\mathrm{CrI}}_{3}$}, Phys. Rev. B 101 (2020) 134418.

\bibitem{DelRe:2021}
L.~Del~Re, A.~Toschi, Dynamical vertex approximation for many-electron systems
  with spontaneously broken su(2) symmetry, Phys. Rev. B 104 (2021) 085120.

\bibitem{Giuliani:2005}
G.~Giuliani, G.~Vignale, Quantum theory of the electron liquid, Cambridge
  university press, 2005.

\bibitem{CINECAlink}
{Description of the ``Galileo100'' HPC cluster at CINECA:
  \url{https://www.hpc.cineca.it/hardware/galileo100}}.

\bibitem{Monkhorst:1976}
H.~Monkhorst, J.~Pack, {Special points for Brillouin-zone integrations}, Phys.
  Rev. B 13 (1976) 5188.

\bibitem{QuantumESPRESSO:Gitlab}
{The latest development version of the \QE{}\, distribution can be downloaded
  from \url{https://gitlab.com/QEF/q-e}.}

\bibitem{QuantumESPRESSO:website}
{The official release of the \QE{}\, distribution can be downloaded from
  \url{https://www.quantum-espresso.org}.}

\end{thebibliography}

\end{document}